\newif\ifsubmode
\submodefalse

\newif\ifprintfig
\printfigtrue

\ifsubmode
  \documentclass[manuscript]{aastex}
  \usepackage{psfig}
  \usepackage{epsf}
  \received{}
  \accepted{}
  \journalid{}{}
  \articleid{}{}
\else
  \documentclass[preprint]{aastex}
  \usepackage{psfig}
  \usepackage{epsf}
  \slugcomment{{\it Astronomical Journal, accepted for publication}}
\fi

\shorttitle{Structure in Brightest Cluster Galaxies}
\shortauthors{Laine et al.}

\newcommand{\etal}{{et al.~}}

\newcommand{\lta}{\lesssim}
\newcommand{\gta}{\gtrsim}

\newcommand{\kms}{\>{\rm km}\,{\rm s}^{-1}}
\newcommand{\kpc}{\>{\rm kpc}}
\newcommand{\Mpc}{\>{\rm Mpc}}
\newcommand{\pc}{\>{\rm pc}}

\newcommand{\Lsun}{\>{\rm L_{\odot}}}

\newcommand{\GG}{\Gamma_{0.05}}

\begin{document}

\title{HUBBLE SPACE TELESCOPE IMAGING OF BRIGHTEST CLUSTER GALAXIES\altaffilmark{1}}

\author{Seppo Laine\altaffilmark{2}, Roeland P.~van der Marel}
\affil{Space Telescope Science Institute, 3700 San Martin Drive,
Baltimore, MD 21218}
\email{laine@stsci.edu, marel@stsci.edu}

\author{Tod R.~Lauer}
\affil{National Optical Astronomy Observatories, P.O. Box 26732, Tucson, AZ
85726}
\email{lauer@noao.edu}

\author{Marc Postman, Christopher P.~O'Dea}
\affil{Space Telescope Science Institute, 3700 San Martin Drive,
Baltimore, MD 21218}
\email{postman@stsci.edu, odea@stsci.edu}

\and

\author{Frazer N.~Owen}
\affil{National Radio Astronomy Observatory, P.O. Box O, Socorro, NM 87801}
\email{fowen@aoc.nrao.edu}


\altaffiltext{1}{Based on observations made with the NASA/ESA Hubble 
Space Telescope, obtained at the Space Telescope Science Institute,
which is operated by the Association of Universities for Research in
Astronomy, Inc., under NASA contract NAS 5-26555. These observations
are associated with proposal \#8683.}
\altaffiltext{2}{Present address: SIRTF Science Center, California Institute
of Technology, 220-6, 1200 East California Boulevard, Pasadena, CA 91125}


\ifsubmode\else
\clearpage\fi


\ifsubmode\else
\baselineskip=14pt
\fi


\begin{abstract} 

We used the {\it HST} WFPC2 to obtain $I$-band images of the centers
of 81 brightest cluster galaxies (BCGs), drawn from a volume-limited
sample of nearby BCGs. The images show a rich variety of morphological
features, including multiple or double nuclei, dust, stellar disks,
point source nuclei, and central surface brightness depressions. High
resolution surface brightness profiles could be inferred for 60
galaxies. Of those, 88\%\ have well-resolved cores. The relationship
between core size and galaxy luminosity for BCGs is indistinguishable
from that of \citeauthor{fab97} (\citeyear{fab97}, hereafter F97) for
galaxies within the same luminosity range. However, the core sizes of
the most luminous BCGs fall below the extrapolation of the F97
relationship $r_{\rm b}\sim L_{V}^{1.15}$. A shallower relationship
$r_{\rm b}\sim L_{V}^{0.72}$ fits both the BCGs and the core galaxies
presented in F97. Twelve percent of the BCG sample lacks a
well-resolved core; all but one of these BCGs have ``power-law'' profiles.
Some of these galaxies have higher luminosities than any power-law
galaxy identified by F97 and have physical upper limits on $r_{\rm b}$
well below the values observed for core galaxies of the same
luminosity. These results support the idea that the central structure
of early-type galaxies is bimodal in its physical properties, but also
suggest that there exist high luminosity galaxies with power-law
profiles (or unusually small cores). The BCGs in the latter category
tend to fall at the low end of the BCG luminosity function and tend to
have low values of the quantity $\alpha$ (the logarithmic slope of the
metric luminosity as a function of radius, at $10 \kpc$). Since
theoretical calculations have shown that the luminosities and $\alpha$
values of BCGs grow with time as a result of accretion, this suggests
a scenario in which elliptical galaxies evolve from power-law profiles
to core profiles through accretion and merging. This is consistent
with theoretical scenarios that invoke the formation of massive black
hole binaries during merger events. More generally, the prevalence of
large cores in the great majority of BCGs, which are likely to have
experienced several generations of galaxy merging, underscores the
role of a mechanism that creates and preserves cores in such merging
events.

\end{abstract}



\clearpage


\section{INTRODUCTION}
\label{s:intro}

Brightest cluster galaxies (BCGs) offer an important probe of the formation of
the central structure of elliptical galaxies, and the role of central massive
black holes in this process. By their very definition, BCGs are highly luminous
elliptical galaxies. The high luminosities of the BCGs, together with their
central location in galaxy clusters, suggest that the processes which shape the
centers of giant ellipticals should be most easily observable in these systems.
Due to their high luminosities, and the strong correlation between galaxy
luminosity and central black hole mass \citep[e.g.,][]{kor01}, BCGs
are expected to harbor the most massive black holes. Due to their central
location in galaxy clusters, we expect them to cannibalize other cluster
galaxies even at the current epoch. The great homogeneity in the global
properties and in the environment of the BCGs gives us perhaps the best
opportunity to understand the central structures found in elliptical galaxies.

Over the past decade, many high resolution imaging studies of elliptical galaxy
centers have been performed with the {\it Hubble Space Telescope}~ 
(e.g., \citealt{cra93,jaf94,bos94,fer94,lau95,byun96,geb96,fab97}, hereafter
F97; \citealt{ver99,quil2000,rav01,rest01}). One of the most interesting
discoveries to emerge from these studies has been that, at {\it HST}
resolution, elliptical galaxies generally have central cusps in their
brightness profiles instead of constant density cores. In addition, elliptical
galaxies display a large range in the central logarithmic slope of their
brightness profile. In luminous ellipticals ($M_{V} \leq -22$) the radial
surface brightness profile shows a clear break in the steepness of the profile
at a resolved radius. Inside the break radius the surface brightness increases
less steeply towards the nucleus, resulting in a shallow cusp at small radii.
These galaxies are commonly called ``core'' galaxies (e.g., F97). No clear
break in the surface brightness profile is seen in less-luminous galaxies
($M_{V} \geq -20.5$), which generally have steep ``power-law'' brightness
profiles into the resolution limit of the {\it HST}.

It is of great interest to understand the origin of the central
density cusps of elliptical galaxies, and their relation to other
galaxy properties.  This is because the cusps are determined by the
same physical processes that shape the formation and evolution of elliptical
galaxies as a whole. Cusps may be created by at least three different
processes. First, violent relaxation in a collisionless system during
galaxy formation is seen to create central density cusps in numerical
simulations \citep*[e.g.,][]{nav97}.  Second, gaseous
dissipation with star formation has been shown to effectively
create cusps, especially during galaxy mergers or accretion
events \citep[e.g.,][]{mih94}. Third, a single massive black
hole in the center of a galaxy may produce a stellar density cusp,
either by the slow growth of a black hole \citep[e.g.,][]{young80} or by its
mere presence \citep{Bah76,Sti98}.

The coexistence of the core and power-law types of central brightness profiles
in the family of elliptical galaxies is intriguing, given the expectation that
less-luminous galaxies are often cannibalized by giant elliptical galaxies.
There is ample evidence that cannibalism is taking place \citep[e.g.,][]{lau98}.
However, the cores of the most luminous galaxies show few signs of the remains
of accreted less-luminous galaxies, whose dense centers would be expected to
arrive at the centers of the giant galaxies largely ``undigested.'' The lack of
evidence for the remains of the centers of accreted galaxies could be due to
the presence of massive black holes, which are believed to be ubiquitous in the
centers of elliptical galaxies \citep[e.g.,][]{Tre02}. A massive black hole at
the center of the brighter galaxy may serve to disrupt the nucleus of the
in-falling, cannibalized galaxy \citep{kelly00,Mer01}. Alternatively, if both
of the merging progenitors contain a central black hole, the merger may
actually be responsible for the creation of a core (F97). A binary black hole
system will eject stars from the nuclear regions by three-body interactions,
thereby lowering the stellar density \citep[e.g.,][]{qui97,mil01}.

Several scenarios for the formation and evolution of the central density cusps
in elliptical galaxies have been compared in detail to {\it HST} data (e.g.,
F97; \citealt{mar99,rav02,Mil02,lau02}). It is clear that the richness of
central structure seen across the entire range of galaxy luminosities and
environments makes it challenging to isolate the physics that is most relevant
to the formation of shallow cusps in luminous galaxies. Study of a sample of
galaxies that is purposely chosen to be homogeneous in global properties may
allow improved insight into the common mechanisms that shape the central
structure of luminous galaxies. This is the approach that we adopt here. We
present the results of an {\it HST} imaging study of a large sample of BCGs. 
We concentrate mostly on the central surface brightness profiles in the present
paper, and report on the correlation between these profiles, black hole masses,
and radio powers, in a later paper.  

\section{SAMPLE AND OBSERVATIONS}
\label{s:sampleobs}

\subsection{Sample}
\label{ss:sample}

Our initial sample consisted of the 119 BCGs in the clusters of the
\citet{abell58} Catalog and its southern extension \citep*{abell89} that
satisfy the following criteria: (i) measured redshift $v$~$\leq$~15,000
km~s$^{-1}$; (ii) galactic latitude $\left| b \right| > 15\degr$; and (iii)
elliptical galaxy morphology. Further slight adjustments to the sample were
made based on uncertain redshifts and the lack of overdensity, as described in
\citet{lau94} and \citet{pos95}. The brightest galaxy in each cluster was found
by looking for the brightest metric magnitude within a given physical radius
\citep{pos95}. We observed the sample in the context of the {\it HST} snapshot
program 8683 (PI: van der Marel). Due to the nature of snapshot programs,
observations were not performed for all 119 galaxies in the sample but only for
a randomly chosen subset of 75. We expanded the observed sample by including
seven BCGs (Abells 262, 569, 1060, 1656, 2162,  3565, and 3742) for which
observations already existed in the {\it HST} Data Archive with the same filter
and camera (three of which were observed previously by two of us and reported
in \citealt{lau98}). The final sample for the present study includes a total of
81 galaxies.\footnote{One of the BCGs (in Abell 3367) for which we obtained
observations has recently been confirmed to be a foreground galaxy with an
observed velocity of 13460~km~s$^{-1}$, compared to the mean velocity of the
cluster at 30477~km~s$^{-1}$ \citep{and98}. At the time the \citet{lau94}
sample was defined, reliable redshifts for all the clusters and BCGs were not
available, and this  galaxy was erroneously determined to be the BCG of Abell
3367. Consequently, we dropped this galaxy from our sample.}

The list of observed galaxies is given in Table~\ref{t:sample} with a
number of basic characteristics. Angular diameter distances were
estimated from the redshifts $z$ after conversion into the cosmic microwave
background (CMB) frame, as described by \citet{lau94}. We used a 
Hubble constant $H_0 = 80 \kms \Mpc^{-1}$ and a cosmology with 
$\Omega = 1$. Because the galaxies are relatively nearby, the dependence 
on the adopted cosmology is negligible for the purposes of the present study. 

An important quantity for the interpretation of the results of our study is the
total luminosity of each galaxy, which is determined by the total observed
magnitude. The metric magnitude of the sample galaxies inside an aperture of
$10 \kpc$ radius has previously been accurately determined. However, BCGs can
be extremely extended and diffuse, and the metric magnitude provides only
limited insight into the total magnitude. We therefore estimated the total
magnitude using the $R$-band surface brightness profiles presented by
\citet{pos95}. We transformed these to the $V$-band using an assumed $V-R =
0.5$, which combines the observed mean $B-R = 1.5$ from \citet{pos95}
with the value $B-V = 1.0$ that is typical for giant  elliptical galaxies
\citep*[e.g.,][]{pel90}. We integrated the best-fitting de
Vaucouleurs $R^{1/4}$ law from the center to infinity, properly taking into
account the galaxy ellipticity, and corrected the result for foreground
Galactic extinction and bandshift (K-correction). The total magnitudes thus
obtained were combined with the luminosity distance (the angular diameter
distance times $[1+z]^2$) to obtain the total absolute magnitude and galaxy
luminosity.

\subsection{Observations and Data Reduction}
\label{ss:obsreduc}

All images were taken with the WFPC2 instrument \citep{Bir01} on board the {\it
HST} between 2000 July 3 and 2001 July 26. The target BCGs were positioned on
the PC chip, which has a pixel size of 0\farcs 0455~$\times$~0\farcs 0455~and a
field of view of 36\farcs 4~$\times$~36\farcs 4.  The placement on the chip was
chosen so as to include also any nearby overlapping cluster galaxies or
multiple nuclei on the PC chip where possible. We used the F814W filter, which
mimics the $I$ band. The total integration time was 1000 seconds, split into
two exposures of 500 seconds to allow for cosmic ray rejection. We employed the
STSDAS task {\tt wfixup} to interpolate (in the x-direction) over bad pixels as
identified in the data quality files. We also used the STSDAS task {\tt
warmpix} to correct consistently warm pixels in the data, using the most recent
warm pixel tables which are provided by the WFPC2 instrument group at STScI
about once a month. The STSDAS task {\tt ccrej} was used to combine the two 500
second exposures. This step corrects most of the pixels affected by cosmic rays
in the combined image. In general, a few cosmic rays remain uncorrected, mostly
when the same pixel was hit in both exposures. Also, a small number of hot
pixels remain uncorrected because they are not listed even in the most recent
warm pixel tables. We corrected these with the IRAF task {\tt cosmicrays},
setting the ``threshold'' and ``fluxratio'' parameters to suitable values that
were selected by a careful comparison of the images before and after correction
to ensure that only questionable pixels were replaced. The photometric
calibration and conversion to the Johnson $I$ band were performed according to
the description given by \citet{hol95}. We corrected for foreground Galactic
extinction using the tables given by \citet{hol95}, assuming a K5 spectrum, and
using the E($B-V$) values from the work of \citet*{schl98}. The K-correction was
made using the values given by \citet*{fuk95}.

\section{IMAGE MORPHOLOGY}
\label{s:images}

Gray-scale images of the central 4\arcsec$\times$4\arcsec~region of the sample
galaxies are shown in Figure~\ref{f:images}. Inspection of the images shows
that all galaxies generally have an elliptical galaxy morphology, consistent
with the ground-based selection criteria. One galaxy, the BCG of Abell 3676,
has a morphology that is somewhat suggestive of a spiral galaxy. This galaxy
was excluded from the discussions of surface brightness profiles in
Sections~\ref{s:sbprofs} and~\ref{s:cusps}. However, even among the galaxies
with unambiguous elliptical galaxy morphologies we find many interesting
morphological features in the images, including multiple nuclei, various dust
absorption features, and embedded stellar disks. We briefly discuss these
before proceeding with a more quantitative analysis.

\subsection{Multiple Nuclei}
\label{ss:multiple}

For all the galaxies in the sample the center can be unambiguously identified.
However, it is not uncommon for BCGs to possess one or more secondary nuclei
separated from the primary galaxy center. A secondary nucleus can either be
physically associated with the BCG, suggesting that a companion galaxy is
currently being accreted, or it can be a mere chance projection of another
cluster member. Morphological studies of individual galaxies have suggested
that the latter possibility is more common \citep{lau88}. Statistical studies
of the kinematics of secondary nuclei have provided support for this
interpretation (\citealt{mer89}, \citeyear{mer91};  \citealt{geb91,Bla92}). In
our sample, 32 of the 81 (40\%) BCGs have at least one secondary nucleus within
the area imaged by the PC chip. These BCGs are identified in the last column of
Table~\ref{t:sample}. We did not make any attempt in the present context to
separate obvious secondary nuclei from probable chance projections.  A study of
the properties of the multiple nuclei at {\it HST} resolution may shed new
light on their nature; however, such an investigation is outside the scope of
the present paper.

In two galaxies, the BCGs of Abell 347 and 3526, it appears that the main
galaxy center itself has a double morphology. High resolution images of the
central regions of these galaxies are shown in Figure~\ref{f:double}. In the
BCG of Abell 347, both brightness peaks have a diffuse nature. We refer the
reader to \citet{lau02} for more discussion on this galaxy. By contrast, in the
BCG of Abell 3526 (the Centaurus Cluster) one of the peaks is unresolved. Such
point source nuclei are generally due to optical emission from an AGN
component, as discussed in Section~\ref{ss:pointnuclei} below. Abell 3526 also
has a spectacular dust lane that wraps around the center (see
Figure~\ref{f:dust}). This has been interpreted as evidence for a recent infall
of a gas-rich galaxy into the BCG \citep*{spa89}. Such an event may also explain
the double nucleus.

\subsection{Dust}
\label{ss:dust}

We visually inspected all the galaxies for signs of absorption by
dust. We did this first in the original images, and subsequently
in images from which an elliptical model for the galaxy light was
subtracted (the construction of these models is described in
Section~\ref{ss:sbanalysis} below). Signs of dust absorption are
evident in 31 of the 81 sample galaxies (38\% of the sample). These
galaxies are identified in the last column of
Table~\ref{t:sample}. The dust can have a variety of different
morphologies. These include nuclear dust disks, dust filaments, patchy
dust, and dust rings or dust spirals around the nuclei. Representative
images of these various dust morphologies are shown in
Figure~\ref{f:dust}. Table~\ref{t:dust} provides a morphological
description of the dust features that we found in the individual
galaxies. The most common classes of dust are dust filaments and
nuclear dust disks. Filamentary dust is found in 14 of the 81
sample galaxies (17\% of the sample) and nuclear dust disks in 11 of 
the 81 sample galaxies (14\% of the sample).

Dust features in elliptical galaxies at {\it HST} resolution have previously
been studied by, e.g., \citet{dok95}, \citet{ver99}, and  \citet{tran01}. 
\citet*{dok95} studied a mixed sample of 64 galaxies from the {\it HST}
archive in the  $V$-band (the F555W filter), and detected dust in about 50\% of
these galaxies. They deduced from the distribution of the axis ratio of the
dust features that 78\% $\pm$ 16\% of early-type galaxies contain nuclear dust.
Our detection rate of 39\% is somewhat lower than theirs. This may reflect the
larger average distance of our sample, yielding a lower spatial resolution in
physical units, and the use of the $V$-band by \citet{dok95},
where the effects of dust are more obvious than in the $I$-band.

\citet{ver99} studied an {\it HST} sample of 19 radio galaxies,
observed through the F555W filter, and detected dust in 17 of them. This large
fraction may be due to a correlation between radio-loudness or radio-power and
the existence of dust. To further address this particular issue we are
observing our sample of BCGs with the VLA at 20 cm. We will report on these
observations, as well as on any possible correlations that we may find with,
e.g., dust properties or nuclear black hole mass, in future papers.

Finally, \citet{tran01} detected dust in 43\% of a distance-limited {\it
HST} sample of 67 early-type galaxies, using images in the $R$-band (F702W
filter), and in 78\% of a sample of 40 galaxies for which they used any optical
images that they could find in the {\it HST} archive. The latter sample was
biased towards detecting dust features by the virtue of relatively high
{\it IRAS} fluxes at 60 and 100 $\mu$m. They found the dust to be in a nuclear
disk in 18\% of the distance-limited sample, and in 38\% of the {\it
IRAS}-biased sample. These numbers compare well with our findings.

\subsection{Nuclear Stellar Disks}
\label{ss:stardisks}

We found that in two of our 81 sample galaxies (2\% of the sample) the
circumnuclear morphology has a high ellipticity, suggesting the possible
presence of an edge-on nuclear disk. These galaxies are identified in the last
column of Table~\ref{t:sample}. This result is consistent with the finding of
nuclear stellar disks in other samples of elliptical galaxies observed with
{\it HST} \citep*[e.g.,][]{lau95,bos98}. Our detection rate is not nearly as
high as that reported by \citet{rest01}, who find evidence for nuclear stellar
disks in 51\% of a distance-limited sample of 67 early-type galaxies \citep[the
same sample was used by][]{tran01}. However, they used ``disky'' perturbations
of the isophotes, characterized by a positive fourth order coefficient of the
cosine term in a Fourier-decomposition of the light profile, as an indicator of
underlying stellar disks. Such a method is likely to provide a much larger
``detected'' fraction of nuclear stellar disks, since positive fourth order
coefficients, however small, will be taken as an indicator of a disk. Our low
detection rate of nuclear stellar disks should also partly reflect the larger
average distance of our sample, compared to the sample of \citet{rest01}. This
results in a lower spatial resolution in physical units, so that small stellar
disks are not resolved. It is also possible that stellar disks are less common
at the high end of the elliptical galaxy luminosity function.

\section{SURFACE BRIGHTNESS PROFILES}
\label{s:sbprofs}

\subsection{Analysis}
\label{ss:sbanalysis}

For quantitative surface brightness profile analysis we ran the images through
20 iterations of the Lucy-Richardson deconvolution routine
\citep{rich72,lucy74}. The number of iterations was decided after considerable
experimentation with varying numbers of iterations. We used 20 iterations here
instead of the 40 used by \citet{lau98} since our data have lower
signal-to-noise ratios ($S/N$). We used a point-spread function (PSF) generated
by the TinyTim software \citep{kri01} for the center of the PC chip of
WFPC2, and a K-type stellar spectrum. The diameter of the synthetic PSF was
3\arcsec, and we tapered the PSF at the edges with an eight pixel Gaussian.

Before performing any fits we inspected the images by eye on the
computer screen. Obvious signs of dust, image defects and foreground
stars were masked. We then fitted ellipses to the isophotes of the
two-dimensional convolved and deconvolved images. The nucleus was
usually found by calculating the centroid in a small box around the
center of the galaxy, but for diffuse cores we used a
cross-correlation technique. After the ellipse fitting, a model was
constructed from the fits and subtracted from the original image. We
then confirmed the dust features by looking at the residual map, and
masked these dust features before performing any profile fitting.

For 13 of the sample galaxies we found that the effects of dust are so severe
that it was not possible to determine a meaningful surface brightness profile.
These galaxies were excluded from the discussion that follows. For another
eight galaxies the effects of dust caused significant uncertainties in the
stellar surface brightness distribution close to the nucleus. Since this is the
region of primary interest in the present context, we excluded these galaxies
from the fitting of the profiles in Section~\ref{ss:paramfits}, and from the
remainder of the discussion. However, we do show the surface brightness
profiles of these galaxies to the extent that they could be determined, in
Figure~\ref{f:profiles}. This leaves a sample of 60 BCGs for which a surface
brightness profile could be determined that is reliable at both small and large
radii. The major-axis surface brightness profiles for these galaxies, together
with  analytical fits discussed in Section~\ref{ss:paramfits}, are shown in
Figure~\ref{f:profiles}. Column (10) of Table~~\ref{t:sample} indicates to
which of the above classes each galaxy belongs. 

Between the nucleus and 0\farcs 5 we used the {\tt profile} task in the VISTA
package to find the surface brightness profile. This task keeps the center
fixed and fits ellipses by sampling the light profile in a circle with a radius
of 1, 2, 3, etc. pixels.  Between the radii of 0\farcs 5 and 1\arcsec~we used
the {\tt snuc} task in VISTA which is capable of fitting ellipses to multiple,
overlapping objects \citep{lau86}. This was required for some of the galaxies
where we saw two or more elliptical nuclei superimposed on the BCG image. For
Abell 347, which has a double-peaked central morphology, we took the profile
within 1\arcsec~from the work of \citet{lau02}, where the profile was extracted
using a one-dimensional cut across the nucleus. Beyond 1\arcsec~we used the
original (not PSF-deconvolved) image for all galaxies, since there is little
gain in deconvolution at large radii, in particular since the $S/N$ is lower
there. We verified the ellipse fits in about half a dozen BCGs by the {\tt
ellipfit} task in the GALPHOT package \citep*{jor92} and the {\tt ellipse} task
in the IRAF package. We found good agreement between the results from the
various packages.

\subsection{Central Point Source Nuclei}
\label{ss:pointnuclei}

We identified 10 BCGs in the sample which have a bright point source in the
very center on top of the smooth stellar surface brightness profile. These
galaxies are identified in the last column of Table~\ref{t:sample}. The point
source component can be identified as an upturn (an inflection point) in the
surface brightness profile at $\sim 0\farcs 1$ from the center; see the panels
for Abell 195, 496, 548, 3526, 3570, 3656, and 3744 in
Figure~\ref{f:profiles}.\footnote{Two other nucleated BCGs (Abell 569 and 2634)
are not shown in Figure~\ref{f:profiles} because of complications in the
determination of their surface brightness profiles due to dust. One other 
nucleated BCG (Abell 2052) is shown in Figure~\ref{f:profiles}, but for this
galaxy the surface brightness profile could only be reliably determined for 
$r~\gta~0\farcs 2$.} Abell 3744 is
listed as ``Nuc?'' in Table~\ref{t:sample} because the dust makes it hard to
establish unambiguously that there is in fact a point source nucleus. Abell
3574 has a bright point source that is offset by $\sim 0\farcs 3$ from the
isophotal center (see Figure~\ref{f:hollow} below). Because the point source is
not at the center, we have not marked this galaxy as ``nucleated'' in
Table~\ref{t:sample}. Of course, its point source could be an off-center
variation to the point-sources seen in the centers of the other galaxies.
However, it could just as well be a foreground star. In the absence of
additional information it is impossible to address the true nature of this
source.

Central point sources in bright elliptical galaxies are generally due to
optical emission from an AGN component. A well-known example is M87
\citep{lau92}, for which the non-thermal nature of the point source has been
confirmed spectroscopically \citep{kor92,mar94}. {\it HST} observations of
samples of radio galaxies have revealed optical point source nuclei in a
majority of the sample galaxies. The detection rates reported by \citet*{chi99}
and \citet{ver02} are 85\% and 57\%, respectively.

\subsection{Parameterized Fits}
\label{ss:paramfits}

To interpret the results we fitted a function of the form 
\begin{equation}
\label{nukerlaw}  
I(r) = I_0 \> (r/r_{\rm b})^{-\gamma} \> 
       (1+ [r/r_{\rm b}]^{\tau})^{\frac{\gamma - \beta}{\tau}}  
\end{equation}  
to the inferred surface brightness profiles. This so-called ``Nuker law''
\citep{lau95,byun96} represents a broken power-law with a turn-over at a break
radius $r_{\rm b}$. The parameter $\tau$ measures the sharpness of the break (it is
usually referred to as $\alpha$, but we use $\tau$ to avoid confusion with
another parameter $\alpha$ that is often used for BCGs; see, e.g.,
\citealt{pos95}). The asymptotic power-law slope is $\gamma$ at small radii and
$\beta$ at large radii. We did not enforce $\gamma$ to be positive in the fit,
but instead allowed both positive and negative values. A negative value of
$\gamma$ corresponds to a surface brightness profile with a central minimum. 
While this may seem counter-intuitive, some galaxies are indeed well described
by such a model (see \citealt{lau02} and Section~\ref{ss:hollow} below). The
quantity $I_0$ determines the normalization of the brightness profile. The
best-fitting Nuker laws are plotted in Figure~\ref{f:profiles} as solid curves.
The parameters of these fits are given in Table~\ref{t:nukerfits}. The fits
were generally performed over the radial range from $0\farcs 02$ (i.e., the
central pixel) to $10\arcsec$ from the galaxy center. For nucleated galaxies only
the data with $r \gta 0\farcs 09$ were included in fit.

The parameters of a NUKER-law fit are well-defined, but some care must be
exercised in their interpretation. For example, $\gamma$ is the logarithmic
slope of the profile for $r \rightarrow 0$. However, whether the observed
profile actually reaches this slope at observationally accessible radii depends
on the values of $r_{\rm b}$ and $\tau$. In the following we will work with the
quantity $\GG$, which we  define to be the power-law slope $-{\rm d}\,\log I /
{\rm d}\,\log r$ at  0\farcs 05 from the galaxy center. This is the last
reliable radius outside the {\it HST} resolution limit in the deconvolved
profiles.  Therefore, it offers the best view of the cusp slope as the radius 
approaches zero. This will correspond to different physical radii in  the BCGs
at varying distances, but in general 0\farcs 05~is $\ll$ $r_{\rm b}$.  Since
the slope may flatten toward the very center, $\GG$ gives us an upper limit to
the asymptotic $\gamma$. $\GG$ is listed for all galaxies  in
Table~\ref{t:nukerfits}.\footnote{For the nucleated galaxies in the sample,
$\GG$ is based on an inward extrapolation of the fit that was performed at
radii $r \gta 0\farcs 09$. Therefore, $\GG$ is somewhat less robustly
established for these galaxies, compared to the remainder of the sample.}
Similar caveats apply to the interpretation of the fit parameter $r_{\rm b}$.
This is the radius at which the fit has its maximum logarithmic curvature
\citep{byun96}. We allowed for $\gamma$ to have negative values to fit the flat
or sometimes downwards sloping central profiles. Note that we restricted the
fitting range to $r~\leq~2\arcsec$~in the BCGs of Abell 76, 347, 634, 3742, and
3747, instead of the usual $r~\leq~10\arcsec$, to better fit the break radii.
In a few cases a good fit was not possible with a Nuker profile, not even after
restricting the fit range. However, such cases were very few (Abell 376, 1177,
and 1314), and do not affect the main results of this paper.

\subsection{BCGs with Central Surface Brightness Depressions}
\label{ss:hollow}

There are six galaxies in the sample for which $\GG < 0$ (i.e., the surface
brightness increases radially outwards at 0\farcs 05), which indicates that
there is a central depression in the surface brightness. These galaxies are the
BCGs of Abell 76, 260, 347, 634, 3574, and 3716 (Figure~\ref{f:hollow}). They
are labeled as `Hollow' in the last column of Table~\ref{t:sample}. One
possible explanation is that the  center of these galaxies may be covered by a
small patch of dust that is not morphologically obvious, and hence was not
masked during the surface brightness profile analysis. Without images in other
passbands it is impossible to assess whether this is the correct explanation.
Such color index images might reveal a subtle reddening towards the center,
indicative of dust absorption. On the other hand, it is quite possible that
there is no dust extinction and that these galaxies do in fact have a
depression in their three-dimensional stellar luminosity density. This has been
argued to be the case for three elliptical galaxies observed in other {\it HST}
programs for which color information is in fact available \citep{lau02}.

\section{CENTRAL CUSP SLOPES}
\label{s:cusps}

\subsection{Core Profiles versus Power-Law Profiles}
\label{ss:corepower}

As discussed in Section~\ref{s:intro}, {\it HST} has been used to study the
surface brightness profiles in various samples of elliptical galaxies. An
important focus of all these studies has been to understand what the central
cusp slopes are, and how this correlates with other galaxy properties. The
Nuker team (\citealt{lau95,byun96,geb96}; F97) found a dichotomy in the
asymptotic power-law slopes at zero radius. The power-law indices were found to
be either larger than $\sim 0.5$ or smaller than $\sim 0.3$. Galaxies with
asymptotic power-law indices $\lta 0.3$ were coined ``core galaxies,'' and
those with power-law indices $\gta 0.5$ were named ``power-law galaxies.'' This
dichotomy in the central power-law slope correlates well with several other
parameters. Core galaxies usually have large total luminosities, boxy central
isophotes, large central velocity dispersions, and low rates of rotation. By
contrast, power-law galaxies usually have smaller total luminosities, disky
isophotes, low central velocity dispersions, and relatively high rates of
rotation. A number of more recent studies have confirmed these results in broad
terms \citep*[e.g.,][]{ver99,quil2000,rav01,rest01}.

It is interesting to see how the results that we have obtained here for BCGs
compare to those obtained previously for other elliptical galaxies. In
Figure~\ref{f:gammavsrad} we plot the central power-law slope $\GG$ versus the
break-radius $r_{\rm b}$ in arcsec. This is similar to figure 3 of F97. We use
this plot, combined with a visual inspection of the brightness profiles in
Figure~\ref{f:profiles}, to distinguish core galaxies from power-law galaxies.
Out of the 60 galaxies for which we have fitted surface brightness profiles,
52 have $\GG \lta 0.3$ and $r_{\rm b} \gta 0\farcs 15$ (rectangular box in
Figure~\ref{f:gammavsrad}). These are core galaxies with well-resolved cores.
Another six galaxies (the BCGs of Abell 189, 261, 419, 912, 1228, 2247) have
$\GG \gta 0.5$. These are power-law galaxies. This leaves two galaxies that
fall in neither of these regions of $(r_{\rm b},\GG)$ space. For these the
classification is more complicated.  One of them (the BCG of Abell 168), while
having $r_{\rm b} < 0\farcs 15$, shows a pronounced turnover to a shallow
slope. We therefore conclude that the  BCG of Abell 168 is a core galaxy. The
other galaxy (the BCG of Abell 1983) has $0.3 \leq \GG
\leq 0.5$ and $r_{\rm b} < 0\farcs 15$. The profile for this galaxy is
steep down to the {\it HST} resolution limit; steeper than core profiles, but
not as steep as power-law profiles. It is possible that in this galaxy we
have just resolved the break in the surface brightness profile, but there are
not enough data points to resolve the smaller cusp slope inside the break
radius.  We classify this galaxy tentatively as `intermediate-slope'
\citep[see also][]{rest01}, and set an upper limit for its core radius.
It is noteworthy that there are no examples of intermediate type galaxies
where $r_{\rm b}$ is well-resolved. We address the significance of the 
power-law vs.~core-type dichotomy in Sections~\ref{ss:globalprop} (correlation
with luminosity) and \ref{ss:hostcor} (correlation with host galaxy
properties), and we discuss the meaning of the results in
Section~\ref{s:disc}. The final classifications for all galaxies are listed in
Table~\ref{t:nukerfits}. The breakdown of the sample is: 53 core galaxies
(88\%), six power-law galaxies (10\%), and one intermediate-slope galaxy
(2\%).

In principle, any core galaxy can be made to look like a power-law
galaxy if it is placed at a sufficiently large distance. It is
therefore important to understand the extent to which profile shape
classifications may depend on distance. F97 discussed
this issue for their sample and found that the distinction between
core and power-law galaxies is an intrinsic one, and is not due to
differences in distance. This is true for our BCG sample as well, for
two reasons. First, there is no correlation between distance and
whether or not a BCG in our sample has a power-law or a core profile;
the average distances are similar for the power-law and the core
galaxies in the sample ($140 \Mpc$ versus $127 \Mpc$, respectively).
Second, the power-law galaxies in our sample have {\it higher} central
surface brightnesses than the core galaxies in our sample. If the
power-law galaxies in our sample were the more distant cousins of the
core galaxies in our sample, seen at distances at which the core is
not resolved, then their observed central surface brightnesses would
be averages over larger physical regions. Since surface brightness
generally falls with radius in a galaxy, the power-law galaxies should then
have had lower observed central surface brightnesses than core galaxies,
contrary to the observations.

As was done by F97, we treat the break radii in power-law (and intermediate)
galaxies differently from the core BCGs. The value of $r_{\rm b}$ for the power-law
galaxies sets a spatial scale by the maximum in the second logarithmic
derivative for a gradually varying profile that is not a pure power-law. 
However, since there is no clear break, these $r_{\rm b}$ values cannot be
meaningfully compared to the break radii observed for  core galaxies. To obtain
the upper limits to the radius of any true break in the power-law galaxies, we
used the following  procedure. For each galaxy we fixed $\gamma$ at 0.3, and
$r_{\rm b}$ at 0\farcs 01, 0\farcs 02,..., respectively.  We then fitted the central
1\arcsec~to optimize the $\tau$, $\beta$, and $I_0$ parameters, and looked for
the $r_{\rm b}$ value at which the $\chi^{2}$ of the fit started to rise
substantially. For the intermediate-type BCG in Abell 1983 we left the previously
fitted $r_{\rm b}$ value (0\farcs 1) as the upper limit. In cases where no clear
minimum in the $\chi^{2}$ value was found, we visually compared the observed
surface brightness profiles to Nuker-law profiles generated with different
$r_{\rm b}$ values, to estimate the upper limit for $r_{\rm b}$. The upper
limits for $r_{\rm b}$ are tabulated in  parenthesis in Table~\ref{t:nukerfits}.

\subsection{Correlations with Galaxy Luminosity}
\label{ss:globalprop}

Figure~\ref{f:gammavsMV} shows the value of $\GG$ for the BCGs vs.~the absolute
galaxy luminosity $M_{V}$. Core profiles ($\GG \leq 0.3$) exist over nearly the
full range of the BCG luminosity function, $-21.8 \geq M_{V} \geq
-25.0$.  However, power-law ($\GG \geq 0.5$) and intermediate-slope ($0.3 \leq
\GG \le 0.5$) profiles exist only in BCGs with relatively low luminosities,
$-21.5 \geq M_{V} \geq -22.6$. Power-law galaxies in the sample studied
by F97 are depicted by the gray region in the figure. Core galaxies in the F97
sample (not shown) fall between the dotted (at $\GG$ = 0.3) and solid ($\GG$ =
0) lines, and between magnitudes $-20.5$ and $-23.5$. F97 summarized their
results by concluding that galaxies with $M_{V} \leq -22$ have core profiles,
galaxies with $M_{V} > -20.5$ have power-law profiles, and galaxies with
intermediate luminosities can have either type of profile. Our results are in
almost perfect agreement with these statements. The only addition is that we
find power-law profiles in galaxies as bright as $M_{V} = -22.6$.

Figure~\ref{f:radvsMV} shows the break radius $r_{\rm b}$ for the BCGs, in
physical units, vs.~the absolute galaxy luminosity $M{_V}$. The area
occupied by core-type galaxies in the sample of F97 is also shown (the
region bracketed by dashed lines). As reported previously by, e.g.,
Kormendy (1985), Lauer (1985), and F97, there is a correlation between
$r_{\rm b}$ and $M_{V}$ in the sense that lower-luminosity galaxies have
smaller break radii. The BCG sample has more core galaxies at high
luminosities than the F97 sample, which has more core galaxies at lower
luminosities. However, in the range of luminosities where they
overlap, the samples display a similar range of $r_{\rm b}$ values. To
quantify this statement we divided the core galaxies in the BCG sample
and the non-BCG core galaxies in the F97 sample in two absolute
magnitudes bins (divided at $M_{V} = -22$). For each magnitude bin we
studied whether the break radius distributions are statistically
equivalent. The Kolmogorov-Smirnov test showed that the $r_{\rm b}$
distributions of BCGs and non-BCG ellipticals are consistent with
being drawn from the same parent population at better than the 99.95\%
level. The Willcoxen signed rank test showed differences between the
two distributions of $r_{\rm b}$ values only at the $0.5 \sigma$ level. So
even though BCGs probably have different accretion histories from
elliptical galaxies in general, this is not reflected in their $r_{\rm b}$
distribution.

While the BCG $r_{\rm b}-L$ relationship does agree with that of F97 over their
common luminosity range, BCGs allow this relationship to be extended to higher
luminosities. The F97 relationship has the form $r_{\rm b}\sim L_{V}^{1.15}$
(see Figure~\ref{f:radvsMV}), but with large scatter --- indeed Lauer (1985)
argued that the scatter seen in the ground-based precursor of this relationship
indicated that cores were more properly described as a multiparameter family.
Figure~\ref{f:radvsMV} shows that the most luminous BCGs fall below the
extrapolation of the F97 relationship. A revised fit over the full luminosity
range of the F97 and present BCG sample gives a flatter relationship $r_{\rm
b}\sim L_V^{0.72}$ (Figure~\ref{f:radvsMVfit}). However, given the large
scatter in $r_{\rm b}$ at all luminosities it is difficult to argue that the
$r_{\rm b}-L$ relationship has really changed form at high luminosities. Also,
it is possible that the most luminous BCGs may be due to events that have
augmented their envelopes, but that have little to do with their central
structure. In this case their $r_{\rm b}$ values are really appropriate to less
luminous galaxies.

At the distance limit of the sample, $0\farcs 05$ corresponds to $\sim 50
\pc$.  This scale is indicated in Figure~\ref{f:radvsMV}. The large majority of
core galaxies are well-resolved at this scale, typically by factors of three or
more (see also Figure~\ref{f:gammavsrad}). By contrast, the upper limits on the
radii of possible breaks in the power-law and intermediate-slope profiles are
generally near the resolution limit. Some of the upper limits are very low,
indicating that any break radius can at most be a few tens of parsecs, a radius
significantly smaller than expected for a core galaxy of a similar luminosity.
This suggests that power-law BCGs are physically different from core-type
galaxies. Support for such an interpretation comes from the observation that
power-law BCGs on average tend to have lower luminosities than core-type BCGs,
and from the clear separation of the power-law galaxies from the horizontal
ridge line of the core galaxies in Figure~\ref{f:gammavsMV}. Similarly,
Figure~\ref{f:gammavsrad} shows that the BCGs outside the core galaxy box are
clearly separated from this box. The failure to find an intermediate-slope BCG
($0.3 < \GG <0.5$) with a well-resolved break shows that such BCGs are rare.
All these considerations provide evidence for a different physical nature of
power-law and core-type BCGs.

The location of NGC 1316 (Fornax A), a peculiar merging galaxy in the F97
sample, is also plotted in Figure~\ref{f:radvsMV}. This galaxy does have a core
\citep{shaya96}, but as F97 emphasized, it is considerably smaller than
those in galaxies of similar luminosity, to the extent that it lies well
outside the F97 $r_{\rm b}-L$ relationship. In Figure~\ref{f:radvsMV} we see
that the core size and luminosity of the BCG in Abell 168 are similar to NGC
1316, while two power-law BCGs, those in Abell 261 and Abell 2247, somewhat
bridge the luminosity gap between the Abell 168 -- NGC~1316 pair, and the
power-law BCGs that conform more closely to the F97 relationship. NGC 1316 thus
appears less as a complete anomaly. Rather, it fits into the bright end of a
class of luminous galaxies that have power-law profiles or cores that are 
substantially smaller than predicted by the F97 $r_{\rm b}-L$ relationship. If
binary black holes are responsible for the larger cores that define the $r_{\rm
b}-L$ relationship, it now becomes interesting to know if any central black
holes in the class of power-law or small-core BCGs have unusually low masses,
or have been ejected altogether, as we will discuss in Section~\ref{s:disc}.

\subsection{Correlations with Other Host Galaxy and Cluster Properties}
\label{ss:hostcor}

It is interesting that we find both core and power-law profiles in our
sample of BCG galaxies, given that the sample is quite homogeneous in
terms of many other properties. To gain some understanding of this
finding, we searched for correlations between the central surface
brightness profile classification and other properties of the BCG or
its host cluster.

There are seven galaxies in the sample that are classified as having
either a power-law profile ($\GG > 0.5$) or an intermediate-slope
profile ($0.3 \leq \GG \leq 0.5$). Four of these galaxies show some
signs of dust in their {\it HST} image (see Table~\ref{t:sample}), i.e.,
57\%. This does not differ significantly from the overall prevalence of
dust in the sample, which is 39\% (Section~\ref{ss:dust}).

One of the seven galaxies with a power-law or intermediate-slope profile
shows morphological evidence for a nuclear stellar disk (see
Table~\ref{t:sample} and Section~\ref{ss:stardisks}). The only other
galaxy in the sample which may have such a nuclear stellar disk has a
core profile. In general, nuclear stellar disks are much more common
in power-law galaxies than in core galaxies \citep{rest01}.  It was
originally suggested that the high central surface brightness of
power-law galaxies was always the result of the presence of stellar
disks seen nearly edge-on \citep{jaf94}. However, this was
refuted by F97 who argued that power-law galaxies have
steeper and higher three-dimensional luminosity densities than core
galaxies, independent of whether or not they harbor a stellar disk.

Six of the seven galaxies with a power-law or intermediate-slope
profile show evidence for multiple nuclei in the {\it HST} image (see
Table~\ref{t:sample}), in the sense defined in
Section~\ref{ss:multiple}. This exceeds the 40\% fraction of the total
BCG sample that show evidence for multiple nuclei. If this were the true
underlying probability of finding multiple nuclei, then the probability
of finding at least six galaxies with multiple nuclei by chance in a
sample of seven is only 2\%. On the other hand, we have not attempted
to carefully discriminate between true secondary nuclei, nearby
cluster members, and background galaxies. We are therefore hesitant to
attach much weight to this statistic.

None of the seven galaxies with a power-law or intermediate-slope profile have
a point-source nucleus (see Table~\ref{t:sample} and
Section~\ref{ss:pointnuclei}). However, only seven of the 53 core galaxies have
such a point-source nucleus, so the lack of power-law BCGs with a point-source
nucleus is not inconsistent with the occurrence fraction in the rest of the
sample. Also, there may be a small systematic effect in the sense that
point-source nuclei are more difficult to identify in power-law galaxies than
in core galaxies.

We have also searched for possible correlations between central surface
brightness profile properties and the larger-scale properties of the
BCGs and their host clusters. The quantities that could potentially be
interesting in this respect are listed in Table~\ref{t:properties}. They
include the following: (a) the absolute $R$-band metric magnitude
$M_{R}$~($10\kpc$) of the BCG inside an aperture of $10 \kpc$ radius;
(b) the parameter $\alpha$, which measures the logarithmic slope of the
metric luminosity as a function of radius, determined at a physical
radius of 10 kpc (Postman \& Lauer 1995); (c) the residual between the
observed metric luminosity and that predicted by the BCG standard-candle
relation between metric luminosity and $\alpha$ \citep{pos95}; (d) the
$B-R$ color; (e) the richness class of the BCG host cluster; (f) the
offset of the BCG from the cluster center in projected position; (g) the
offset of the BCG from the cluster center in line-of-sight velocity; (h)
the morphological classification of the cluster; (i) the velocity
dispersion of the cluster; and (j) the X-ray luminosity of the cluster.
We checked for correlations between each of these quantities and the
central surface brightness profile classifications obtained from the
{\it HST} data (Table~\ref{t:nukerfits}). We found a meaningful
correlation with only two quantities: $M_{R}$~($10\kpc$) and $\alpha$.
We did not test for a correlation with the cluster  elliptical/spiral
ratio, because this quantity is not readily available for most of the
clusters in our sample. However, this ratio is known to correlate with
the cluster morphological type \citep[e.g.,][]{sara88}. Since there is 
no correlation with the latter, we do not suspect the existence of a 
correlation between the nuclear cusp slope of a BCG and the  
elliptical/spiral ratio of the host cluster.

It is no great surprise that there is a correlation between central
surface brightness profile classification and metric luminosity
$M_{R}$~($10 \kpc$). After all, we know that there is a correlation with
total luminosity (Figure~\ref{f:gammavsMV}). We find the
correlation with metric luminosity to be very similar. The power-law
BCGs are all at the low-end of the BCG metric luminosity function. Only Abell
261 has an absolute metric $R$-band magnitude that is (somewhat)
brighter than the sample mean (which is $M_{R}$ [10 kpc] = $-22.47$).

The correlation between central surface brightness profile
classification and the parameter $\alpha$ is also not entirely
unexpected, because $\alpha$ is itself strongly correlated with metric
luminosity \citep{pos95}. Figure~\ref{f:gammavsalpha} shows
$\GG$ versus $\alpha$. The seven BCGs that are classified as power-law
or intermediate-slope all have relatively small $\alpha$ values. The
parameter $\alpha$ depends on the slope of the intensity profile at
$10 \kpc$, which corresponds to $10\arcsec$ at the distance limit of our
sample. This exceeds the scale of $0\farcs 05$ at which $\GG$ is
measured by a factor of $200$. It also exceeds the break radius of the
core galaxies in our BCG sample by a factor of five or more. The
correlation between $\GG$ and $\alpha$ therefore has true physical
meaning, and is not merely a tautology.

\section{DISCUSSION}
\label{s:disc}

Previous studies with {\it HST} demonstrated that very bright elliptical
galaxies almost always have core-type brightness profiles. BCGs are by
definition the brightest galaxies in their host clusters, and as a class are
known as the brightest galaxies in the universe. The {\it a priori} expectation
for our study was therefore that we would predominantly find core galaxies
amongst the BCGs. Indeed, the observations show that core-type profiles exist
in 88\% of the sample. This finding is in itself quite important. It shows that
cores are dominant even in the highest luminosity galaxies, which, by the
virtue of their central position in galaxy clusters, accrete significantly even
at the present epoch. One can also turn this argument around, and argue that it
is actually surprising that we have identified BCGs without core-type profiles.
However, our results are not in contradiction with the trends that have been
established previously for samples that were more heavily weighted towards
lower luminosity galaxies. The BCGs with power-law and intermediate-slope
profiles all reside at the low end of the BCG luminosity function, as would
have been expected on the basis of previous work. In fact, the BCGs fit almost
seamlessly into the previously established trends; see in particular
Figure~\ref{f:gammavsMV}. The only novelty is that we find that power-law
profiles can occur at somewhat brighter magnitudes than was previously found,
up to $M_{V}$ = -22.6.

Galaxies in general are thought to form through hierarchical accretion of
smaller subunits. This is true in particular for BCGs, which live in
environments where the continued infall of smaller subunits is common, even at
the present epoch. Semi-analytical models of galaxy formation make explicit
predictions for the merging histories of galaxies, as a function of the
circular velocity $V_{\rm circ}$ of the halo \citep*[e.g.,][]{kau01}. BCGs have
$V_{\rm circ} \approx 400$--$500 \kms$ \citep{ger01}. For BCGs this
implies that the average time since the last accretion event with a mass ratio
larger than 1:10 is 40\% of the Hubble time. Only $\sim 10$\% of BCGs are
expected to not have had such an accretion event for the last 2/3 of a Hubble
time. From an observational perspective, it is well known that BCGs are
generally more luminous and more extended than elliptical galaxies found in
other environments \citep[e.g.,][]{ton87}. This has been attributed directly to
accretion events \citep{lau88}. \citet{hau78} performed
semi-analytical calculations of the evolution of a BCG through accretion
events. As time progresses, the BCG becomes more luminous. At the same time,
the radial surface brightness gradients become smaller, which causes $\alpha$
to increase. The most natural interpretation of the observed spread in
luminosities and $\alpha$ values for BCGs is therefore that different BCGs have
undergone different amounts of accretion. In this view, the BCGs with the
smallest luminosities and the smallest $\alpha$ values are the ones that have
had the least pronounced accretion history.

We have found that the central surface brightness profile shapes of BCGs are
strongly correlated with both their luminosities (Figure~\ref{f:gammavsMV}) and
$\alpha$ values (Figure~\ref{f:gammavsalpha}). We interpret this as direct
evidence that the central surface brightness profile shapes of elliptical
galaxies are related to their accretion and merging history. In particular, our
results support scenarios in which elliptical galaxies evolve from power-law
profiles to core profiles through accretion and merging. This is consistent
with theoretical investigations of the formation of core galaxies (e.g., F97,
\citealt{Mer01,mil01}). In these studies most galaxies start out with steep
power-law surface brightness profiles (presumed to be due to any of the
physical processes discussed in Section~\ref{s:intro}) and a central black
hole. As two galaxies merge, the formation and coalescence of a binary black
hole system creates a core-type surface brightness profile \citep{qui97}. More
massive galaxies have undergone more accretion events \citep[e.g.,][]{kau01},
and this causes core-type profiles to be more common amongst the most luminous
galaxies. This scenario also receives support from other arguments 
\citep*[e.g.,][]{rav02,Mil02}. The fact that some high-luminosity galaxies
have power-law profiles is not inconsistent with this scenario. It might be
that these galaxies simply have not had a significant accretion event.
Merger-tree predictions, such as those presented by Kauffmann \etal (2001), do
not rule this out. 

Another possibility is that cannibalism has occured during a period when a
massive black hole is absent in the center of the high-luminosity accreting
galaxy.  The initial generation of a core as a merger endpoint appears to
require the formation of a binary black hole as the nuclei of the two merging
galaxies are brought together. If the binary later hardens to the point where
its orbital velocities exceed the escape velocity of the merged galaxy, then
later accretion of a third black hole may lead to an interaction in which all
three black holes are ejected. Any dense nucleus cannibalized after this event
might then settle into the core of the luminous galaxy without being disrupted.
\citep*{vhm02} have studied the central accretion of massive black holes during
the formation of luminous galaxies by hierarchical merging, concluding that in
the early stages of galaxy formation, up to 8\%\ of the three-body black hole
interactions may result in all black holes being ejected from the system. The
present small fraction of BCGs with anomalously small cores may be consistent
with this scenario.

While the scenario outlined above naturally explains the existence of core-type
profiles in the highest luminosity galaxies, it is less successful in answering
another related question: why do we not observe core-type profiles in some
galaxies with relatively low luminosities?  After all, some of these must have
had merging and accretion events too. This suggests that there are mechanisms
at work in mergers that prevent the formation of a core under certain
circumstances. One possible explanation is that mergers in low-luminosity
galaxies may involve more gaseous dissipation \citep[e.g.,][]{mih94}. The
resulting star formation could mask the influence of the black hole binary.
There is some evidence for this from the fact that low-luminosity ellipticals
tend to be more disky and rotate more rapidly than high-luminosity ellipticals.
This is consistent with the hypothesis that dissipation has been important in
their formation (e.g., F97). Future color or line strength information would be
very useful for testing this scenario by probing the age of the stars in the
nucleus. Other possible explanations for the complete absence of core-type
profiles in low luminosity galaxies include (a) the survival of the stellar
nucleus of a small accreted companion, which would then fill in any core that
would otherwise have been created by the black hole binary; and (b) the failure
of an accreted companion to make its way all the way to the center, so that a
black hole binary never forms. However, the most recent calculations do not
favor these scenarios, unless they take place under certain special
circumstances \citep{kelly99,kelly00,Mer01}. On the other hand, if these
scenarios were correct, it would be unclear why the same processes would not
prevent the formation of cores in most of the high luminosity galaxies.

\section{CONCLUSIONS}
\label{s:conc}

We have presented a study of an unbiased sample of 81 nearby BCGs with
an elliptical galaxy morphology. We observed these systems in the
$I$-band with the {\it HST} WFPC2 camera. The images show a rich
variety of morphological structures: 32 galaxies (40\% of the sample)
have multiple nuclei or nearby companions; two galaxies (2\%) have a
double morphology at the sub-arcsec scale; 31 galaxies (38\% of the
sample) show morphological evidence for dust in the form of dust
disks, filaments, patches, rings or spirals; two galaxies (2\%) have a
circumnuclear morphology that has a high ellipticity, suggesting the
possible presence of an edge-on nuclear disk; ten galaxies (12\%) have
a point-source nucleus on top of the smooth stellar surface brightness
profile, presumably due to non-thermal emission from an AGN; and six
galaxies (7\%) have a central depression in their surface brightness
distribution, possibly due to an actual decrease in the
three-dimensional stellar luminosity density.

With the help of isophotal fitting of PSF-deconvolved images we
determined reliable surface brightness profiles for 60 of the sample
galaxies. The profiles were fitted with the so-called Nuker-law
parameterization. Following previous authors, we classified the
surface brightness profiles on the basis of the fit parameters and
visual inspection of the profiles. We identified 53 galaxies (88\% of
the sample) as ``core'' galaxies (these have a shallow cusp and a
well-defined break), six galaxies (10\%) as ``power-law'' galaxies
(these have a steep brightness profile all the way into the center),
and one galaxy (2\%) that has an ``intermediate slope.'' We have
studied how the Nuker-law fit parameters and the profile
classifications relate to the global properties of the BCGs and their
host clusters. We have also compared the results to those obtained
previously for other samples of elliptical galaxies, which were more
heavily weighted towards low-luminosity objects.

Previous studies with {\it HST} demonstrated that bright elliptical galaxies
have shallow core-type profiles. Our study underscores this result by showing
that cores are dominant even in the highest luminosity galaxies, which by the
virtue of their central position in galaxy clusters accrete significantly even
at the present epoch. The finding that 12\% of the BCG sample has a power-law
or intermediate-slope profile is not in contradiction with earlier work. The
power-law and intermediate-slope BCGs have $-21.5 \geq M_{V} \geq -22.6$, which
puts them at the low-luminosity end of the BCG luminosity function (which
itself extends to $M_V = -25.0$). Our results differ from the general rules
obtained by F97 only in the sense that we find that power-law profiles can
occur in galaxies as bright as $M_{V} = -22.6$. F97 quoted $M_{V} = -22.0$ as the
upper limit for the absolute $V$-magnitude of the power-law galaxies.

We determined upper limits on the sizes of any potential cores in the power-law
and intermediate-slope galaxies in the BCG sample and find these to be much
smaller than typical core sizes in core-type galaxies. Combined with the lower
luminosity of power-law galaxies and the paucity of intermediate-slope galaxies,
our findings are consistent with the hypothesis that the power-law galaxies are
physically different from core-type galaxies. The type of the central profile
(core vs. power-law) does not appear to correlate with galaxy distance or
morphology, the position or velocity offset from the cluster center, or the 
velocity dispersion, X-ray luminosity, morphology, or richness class of the 
host cluster. However, there is a significant correlation with the quantity $\alpha$
that is used in the standard-candle relations for BCGs. It measures the
logarithmic slope of the metric luminosity as a function of radius, at a fixed
physical radius (here chosen to be $10 \kpc$). BCGs with power-law or
intermediate-slope profiles all have relatively low values of $\alpha$. The
connection of the BCG properties, including their central black hole masses, to
the BCG radio luminosities and cluster cooling flows will be addressed in a 
future paper.

These results for BCGs provide important new insight into the physical
processes that shape the centers of elliptical galaxies. We emphasize that the
BCGs form a very homogeneous population of galaxies that presumably had
very similar formation and evolutionary histories. Given their positions in the
centers of prominent clusters, accretion and merging must have played an
important part in their evolution. Growth by accretion and merging provides a
natural explanation for their unusually extended envelopes. Theoretical
calculations have shown that the luminosities and $\alpha$ values of BCGs grow
as a result of accretion. This suggests that BCGs with low luminosities and low
$\alpha$ values have had the least pronounced accretion histories. Our
observations show that these are the galaxies that are most likely to have
power-law profiles. We interpret this as evidence that elliptical galaxies
evolve from power-law profiles to core-profiles through accretion and merging.
Such a morphological evolution can be naturally explained by theoretical
studies where the merging progenitors both have central black holes. After the
progenitor galaxies have merged, the black holes form a central binary that
ejects stars from the central region, giving rise to a central core with a
shallow cusp. The massive black holes serve to both create and protect cores in
the centers of luminous ellipticals.


\acknowledgments 
Support for proposal \#8683 was provided by NASA through a grant from
the Space Telescope Science Institute, which is operated by the
Association of Universities for Research in Astronomy, Inc., under
NASA contract NAS 5-26555. This research has made use of the
NASA/IPAC Extragalactic Database (NED) which is operated by the Jet
Propulsion Laboratory, California Institute of Technology, under
contract with the National Aeronautics and Space Administration.

\clearpage




\ifsubmode\else
\baselineskip=10pt
\fi


\clearpage


\ifsubmode\else
\baselineskip=14pt
\fi


\newcommand{\figcapimages}{Gray-scale images of the 81 BCGs in our sample.  The
gray-scale is arbitrary, adjusted to show the cores and central dust features
as well as possible. The images are shown before any deconvolution was applied.
The images are shown as positives (bright areas are shown with light colors).
Only the central 4\arcsec $\times$ 4\arcsec~is shown to emphasize the core
structure.  The direction to north in the images is listed in
Table~\ref{t:sample}.\label{f:images}}

\newcommand{\figcapdouble}{Gray-scale images of the central regions of 
the two galaxies in our sample which have a double morphology in the
central arcsec. The images are shown as negatives (bright areas are
shown as dark). The gray-scale was manually adjusted for each galaxy
to achieve the best contrast. The axes are labeled in
arcsec.\label{f:double}}

\newcommand{\figcapdust}{Gray-scale images showing examples of the  different
morphological dust structures observed in our sample galaxies. All images,
except for that of the BCG of Abell 3526, have been high-pass filtered to bring
the dust features out more clearly. The gray-scale is linear but the levels are
arbitrary. The images are shown as positives (dust is shown with  darker
shades). The axes are labeled in arcsec.\label{f:dust}}

\newcommand{\figcapprofiles}{Major axis surface brightness profiles of the 68
galaxies in the sample for which this profile could be  determined, as
described in Section~\ref{ss:sbanalysis}. Solid curves show the best fits of a
Nuker-law, as parameterized by equation~(\ref{nukerlaw}). The parameters of
these fits are discussed in the text and are listed in Table~\ref{t:nukerfits}.
The figure includes eight galaxies for which the effects of dust precluded us
from performing a reliable fit to the observed profile. The bottom axis of each
plot is in units of arcsec, running from 0\farcs 01~to 11\arcsec. The top axis
is labeled in physical units of parsecs. The break radius for the
core-type BCGs is shown with an upward pointing arrow near the top
axis. The ordinate shows the calibrated $I$-band surface brightness after
correction for Galactic foreground extinction and bandshift
(K-correction).\label{f:profiles}}

\newcommand{\figcaphollow}{Gray-scale images of the central regions of the
galaxies in our sample with a central light depression. The image of the sixth
galaxy with a central depression, Abell 347, is shown in Figure~\ref{f:double}.
The images are shown as negatives (bright areas are shown as dark). The
gray-scale was manually adjusted for each galaxy to achieve the best contrast.
The axes are labeled in arcsec. Abell 3574 has a bright unresolved source at
$0\farcs 3$ from the isophotal center (the apparent extent in the image is due
only to the adopted contrast). This could be a foreground star or an off-center
AGN (see~Section~\ref{ss:pointnuclei}). It was masked in the surface brightness
profile analysis.\label{f:hollow}}

\newcommand{\figcapgammavsrad}{Power-law slope $\GG$ of the surface
brightness profile at $r=0\farcs 05$ versus the observed break radius $r_{\rm
b}$ in arcsec. BCGs with some dust (see Section~\ref{ss:dust}) are shown as
open triangles; BCGs with a nuclear point source, presumably due to an AGN (see
Section~\ref{ss:pointnuclei}), are shown as open circles; BCGs with hollow
centers (see Section~\ref{ss:hollow}) are shown as circles with an enclosed
plus sign; the remaining BCGs are shown as open squares. The solid horizontal
line indicates $\GG = 0$. Galaxies below this line  have a central depression
in their surface brightness  (see Section~\ref{ss:hollow}). The dashed line
indicates $\GG = 0.5$.  Galaxies above this line are classified as
``power-law'' galaxies. The dotted rectangular box indicates the region of
parameter space with $\GG \lta 0.3$ and $r_{\rm b} \gta 0\farcs 15$. Galaxies in this
region are classified as ``core'' galaxies.\label{f:gammavsrad}}

\newcommand{\figcapgammavsMV}{Power-law slope $\GG$ of the surface
brightness profile at $r=0\farcs 05$ versus the total absolute $V$-band galaxy
magnitude $M_{V}$. The symbols are the same as in Figure~\ref{f:gammavsrad}.
The solid line indicates $\GG = 0$.  Galaxies below this line have a central
depression in their surface  brightness (see Section~\ref{ss:hollow}). The
dotted line indicates $\GG = 0.3$. Galaxies below this line are (generally)
core galaxies. Galaxies above this line are classified either as power-law
($\GG > 0.5$) or intermediate-slope ($0.3 \leq \GG \leq 0.5$). The gray region
depicts the area occupied by the power-law and intermediate type galaxies in
F97. For comparison, the core-type galaxies in the sample of F97 (not shown)
occupy the magnitude range $-20.5 \geq M_{V}$ $\geq -23.5$ and have $0 \leq
\GG \leq 0.3$. \label{f:gammavsMV}}

\newcommand{\figcapradvsMV}{Break radius $r_{\rm b}$ of the  surface
brightness profile in parsecs versus the total absolute $V$-band galaxy
magnitude $M_{V}$. The 53 BCG galaxies with core-type profiles listed in
Table~\ref{t:nukerfits} are plotted with open symbols as in
Figure~\ref{f:gammavsrad}. Upper limits for power-law galaxies, calculated from
the values shown in parenthesis in Table~\ref{t:nukerfits} are shown with
filled symbols and downward pointing arrows. The dash-dotted horizontal line
indicates the  physical scale of $50 \pc$ that spans $0\farcs 05$ at the
distance  limit of our BCG sample. The large majority of BCG core galaxies are 
resolved at this scale by a factor of three or more. A linear fit to  the
correlation displayed by the F97 core galaxies is shown with a solid  line,
surrounded by two parallel dashed lines showing the lower and upper  limits for
the spread of F97 core galaxies. The vertical dash-dotted line  shows the upper
magnitude limit for the F97 power-law galaxies. The position  of Fornax A, a
peculiar galaxy in the sample of F97, is also plotted. The BCGs with the
smallest physical core radii have been labeled, together with Abell 168, the
core-type galaxy with the smallest physical radius of the core.\label{f:radvsMV}}

\newcommand{\figcapradvsMVfit}{Break radius $r_{\rm b}$ for the core-type
galaxies in the combined BCG and F97 samples, plotted versus the total absolute
$V$-band galaxy magnitude $M_{V}$. The core galaxies from the F97 sample are
also shown. The plot symbols are the same as before in Figs.~\ref{f:gammavsrad}
-- \ref{f:radvsMV}; the F97 core galaxies are shown with filled squares. The
best linear fit is shown with a solid line.\label{f:radvsMVfit}}

\newcommand{\figcapgammavsalpha}{Power-law slope $\GG$ of the
surface brightness profile at $r=0\farcs 05$ versus the $\alpha$
parameter. The latter measures the logarithmic slope of the metric
luminosity as a function of radius, determined at a physical radius of
10 kpc \citep{pos95}. The 60 BCG galaxies listed in
Table~\ref{t:nukerfits} are shown as open symbols (as in
Figure~\ref{f:gammavsrad}). The solid line indicates $\GG =
0$. Galaxies below this line have a central depression in their
surface brightness (see Section~\ref{ss:hollow}). The dotted line in
both panels indicates $\GG = 0.3$. Galaxies below this line are
(generally) core galaxies.  Galaxies above this line are classified
either as power-law ($\GG > 0.5$) or intermediate-slope ($0.3 \leq \GG
\leq 0.5$).\label{f:gammavsalpha}}


\ifsubmode
\figcaption{\figcapimages}
\figcaption{\figcapdouble}
\figcaption{\figcapdust}
\figcaption{\figcapprofiles}
\figcaption{\figcaphollow}
\figcaption{\figcapgammavsrad}
\figcaption{\figcapgammavsMV}
\figcaption{\figcapradvsMV}
\figcaption{\figcapradvsMVfit}
\figcaption{\figcapgammavsalpha}
\clearpage
\else\printfigtrue\fi

\ifprintfig


\clearpage
\begin{figure}
\epsfxsize=0.8\hsize
\centerline{\epsfbox{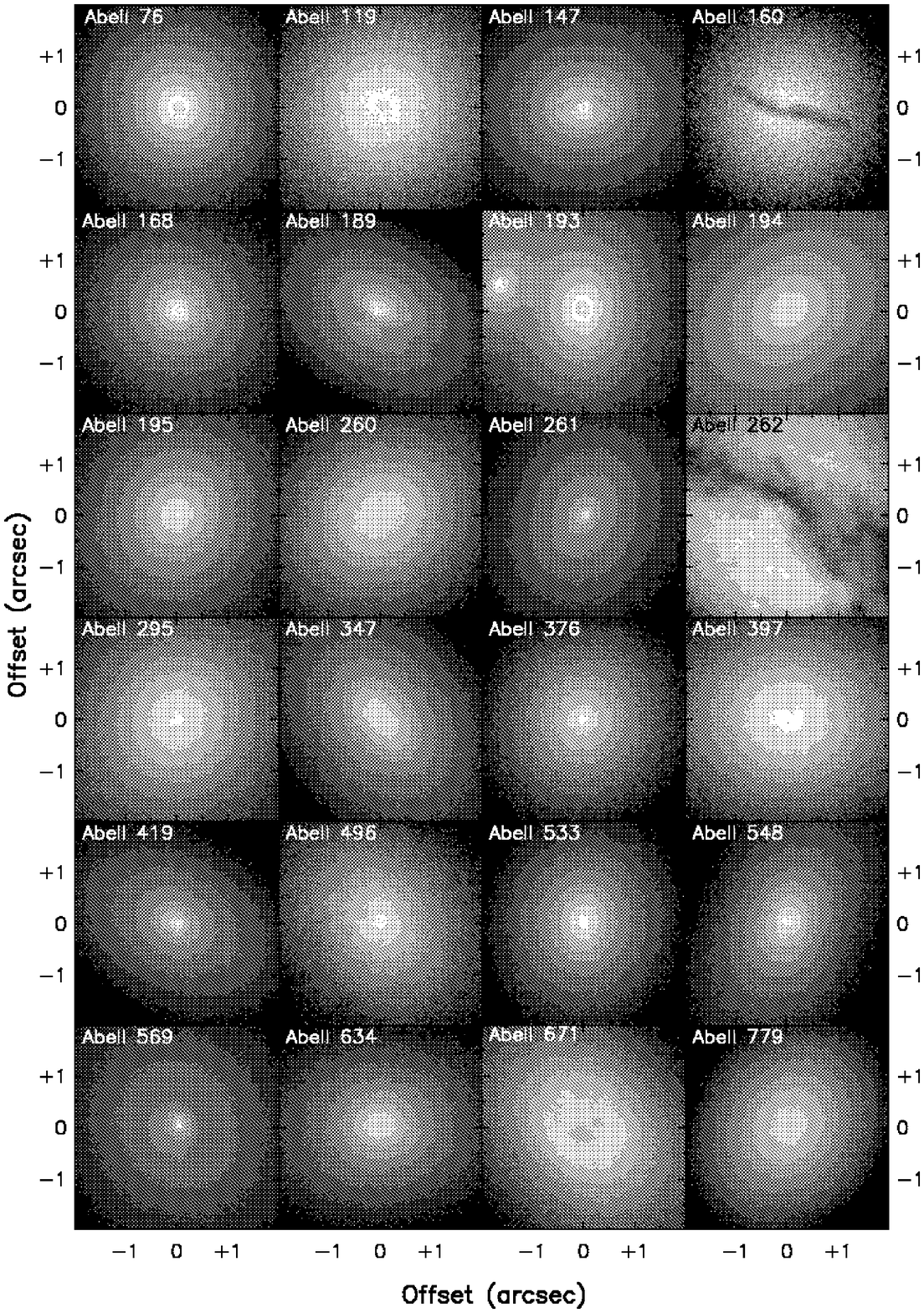}}
\ifsubmode
\vskip3.0truecm
\setcounter{figure}{1}
\centerline{Figure~\thefigure}
\else\figcaption{\figcapimages}\fi
\end{figure}


\clearpage
\begin{figure}
\epsfxsize=0.9\hsize
\centerline{\epsfbox{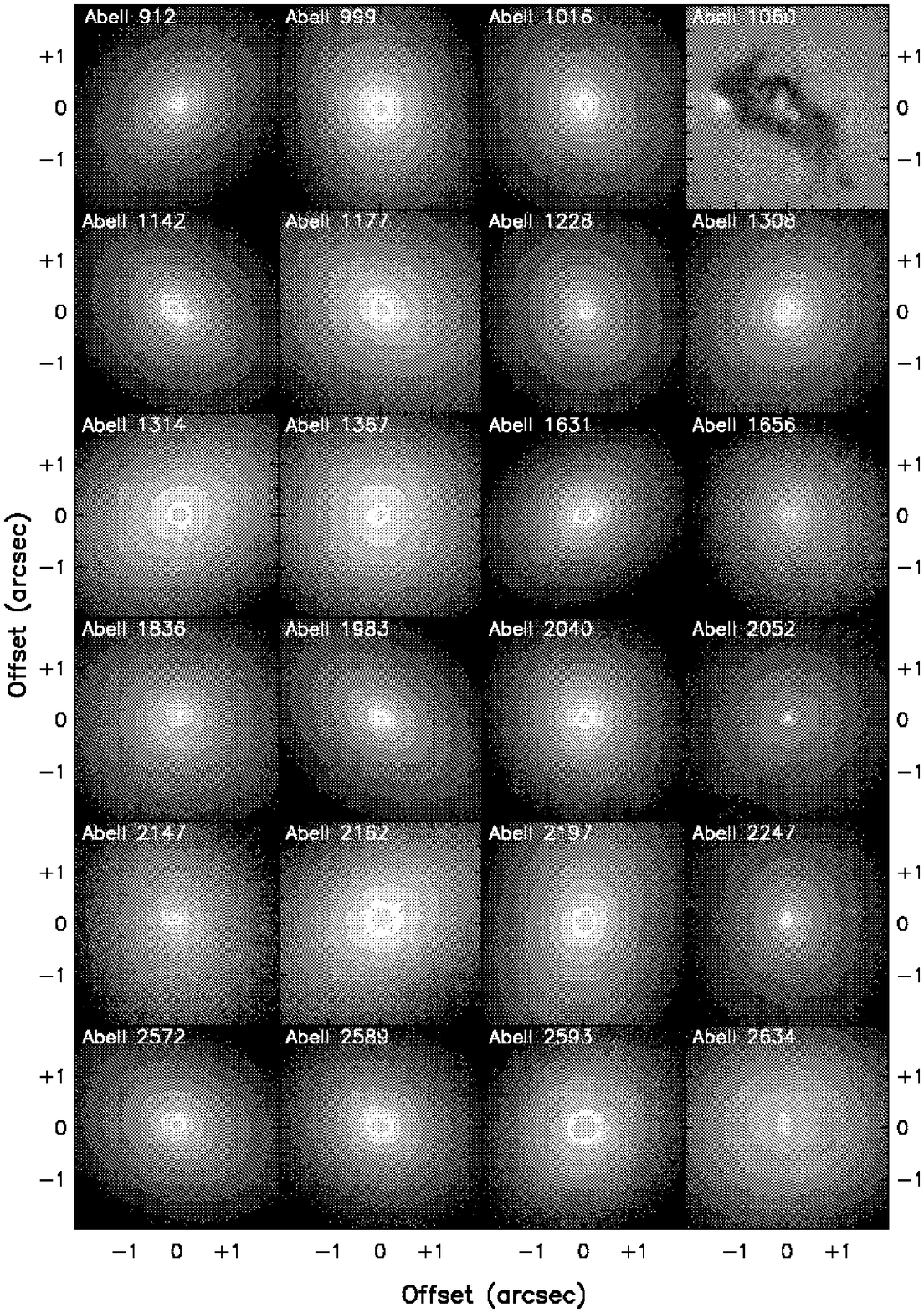}}
\ifsubmode
\vskip3.0truecm
\centerline{Figure~\thefigure\ (continued)}
\else
\centerline{Fig. \thefigure.--- (continued)}
\fi
\end{figure}


\clearpage
\begin{figure}
\epsfxsize=0.9\hsize
\centerline{\epsfbox{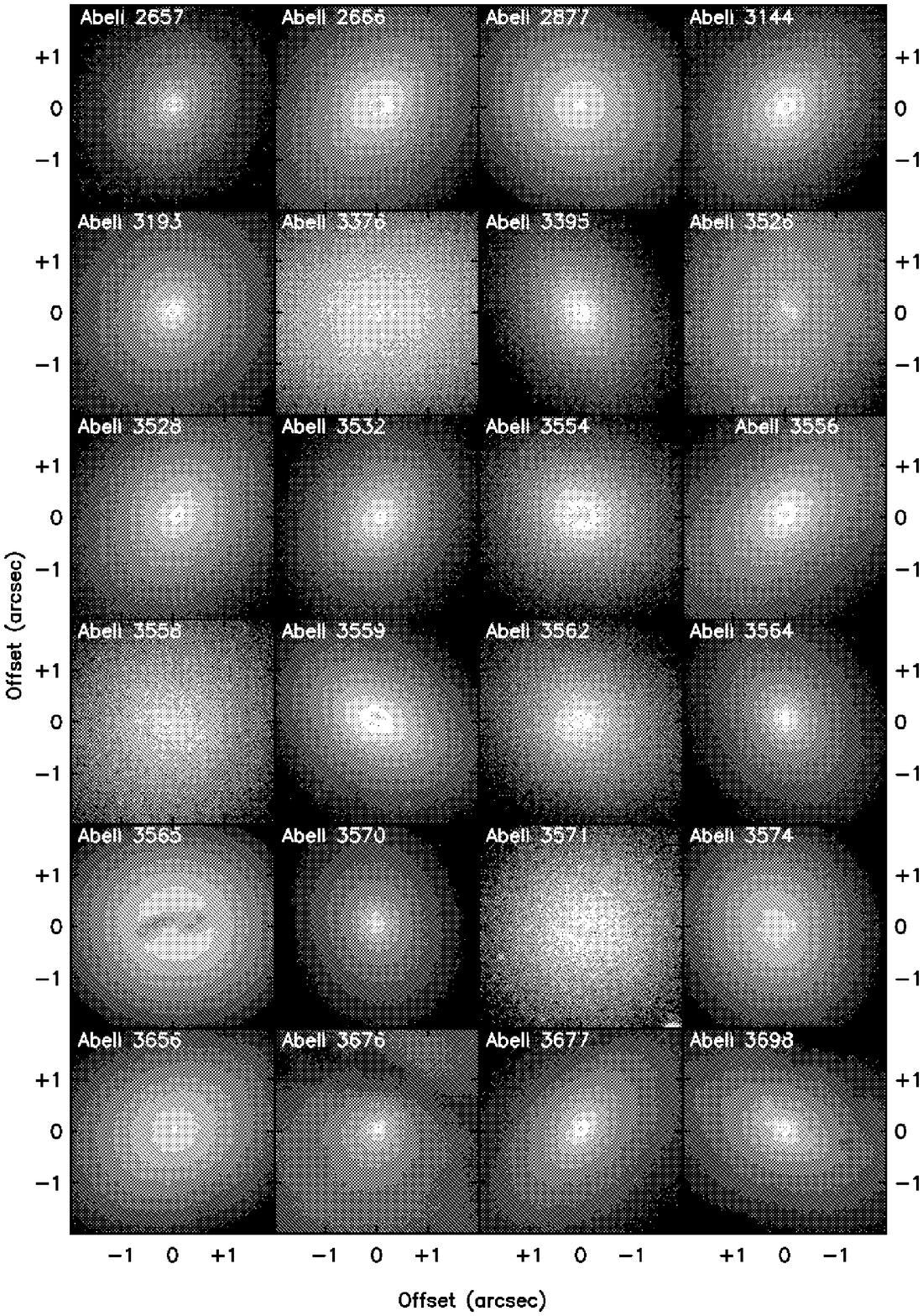}}
\ifsubmode
\vskip3.0truecm
\centerline{Figure~\thefigure\ (continued)}
\else
\centerline{Fig. \thefigure.--- (continued)}
\fi
\end{figure}


\clearpage
\begin{figure}
\epsfxsize=0.9\hsize
\centerline{\epsfbox{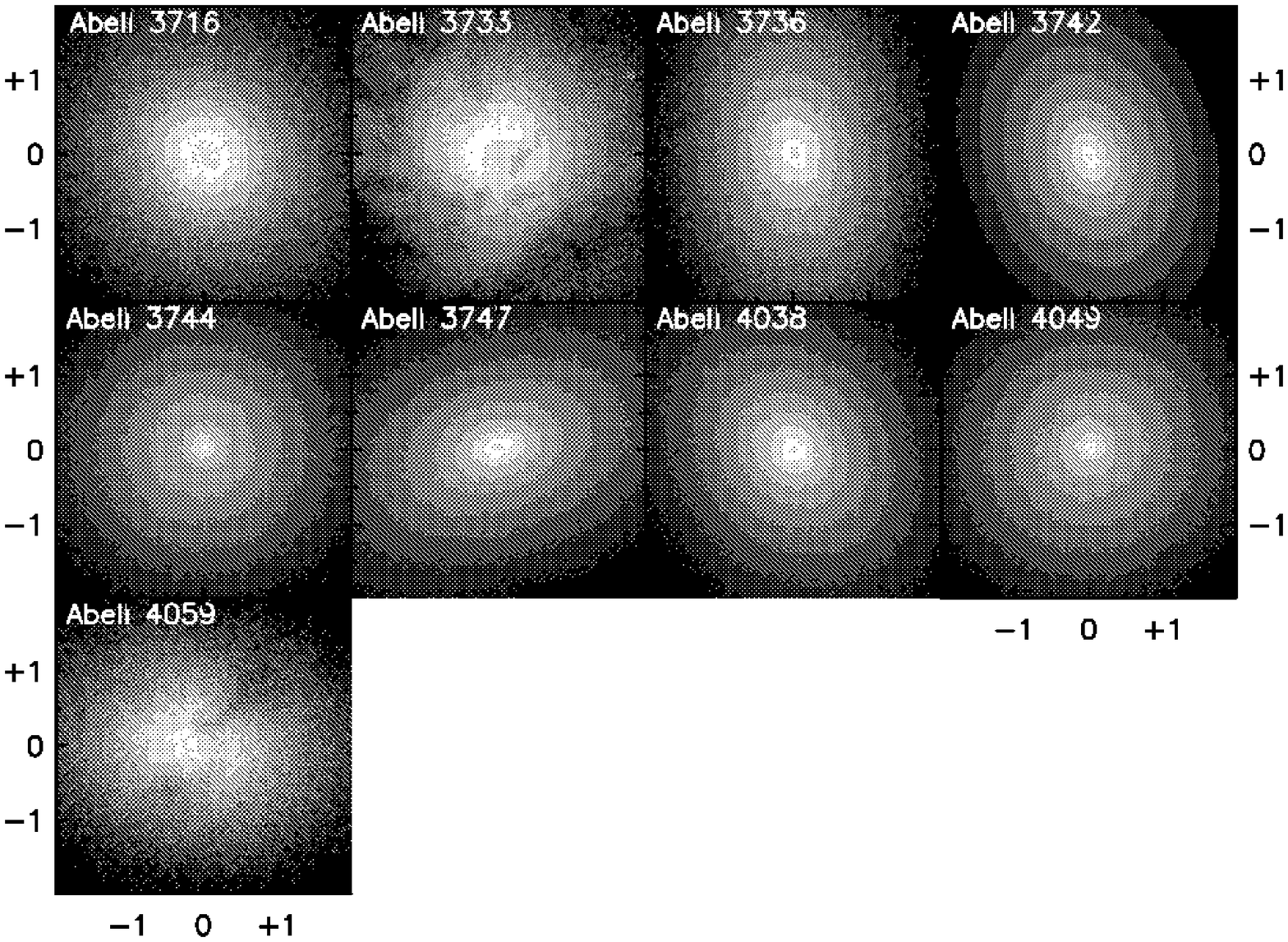}}
\ifsubmode
\vskip3.0truecm
\centerline{Figure~\thefigure\ (continued)}
\else
\centerline{Fig. \thefigure.--- (continued)}
\fi
\end{figure}


\begin{figure}
\epsfxsize=0.7\hsize
\centerline{\epsfbox{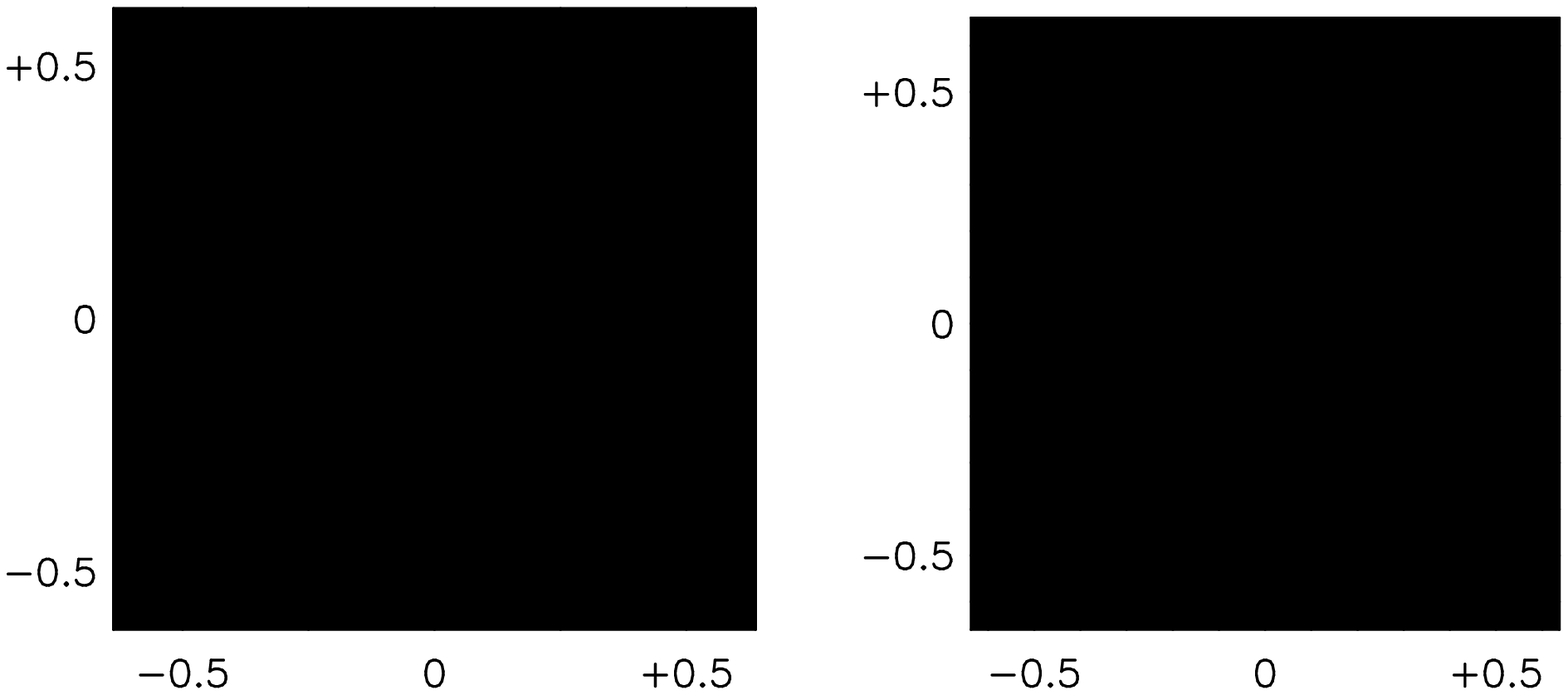}}
\ifsubmode
\addtocounter{figure}{1}
\vskip3.0truecm
\centerline{Figure~\thefigure}
\else\figcaption{\figcapdouble}\fi
\end{figure}


\begin{figure}
\epsfxsize=0.9\hsize
\centerline{\epsfbox{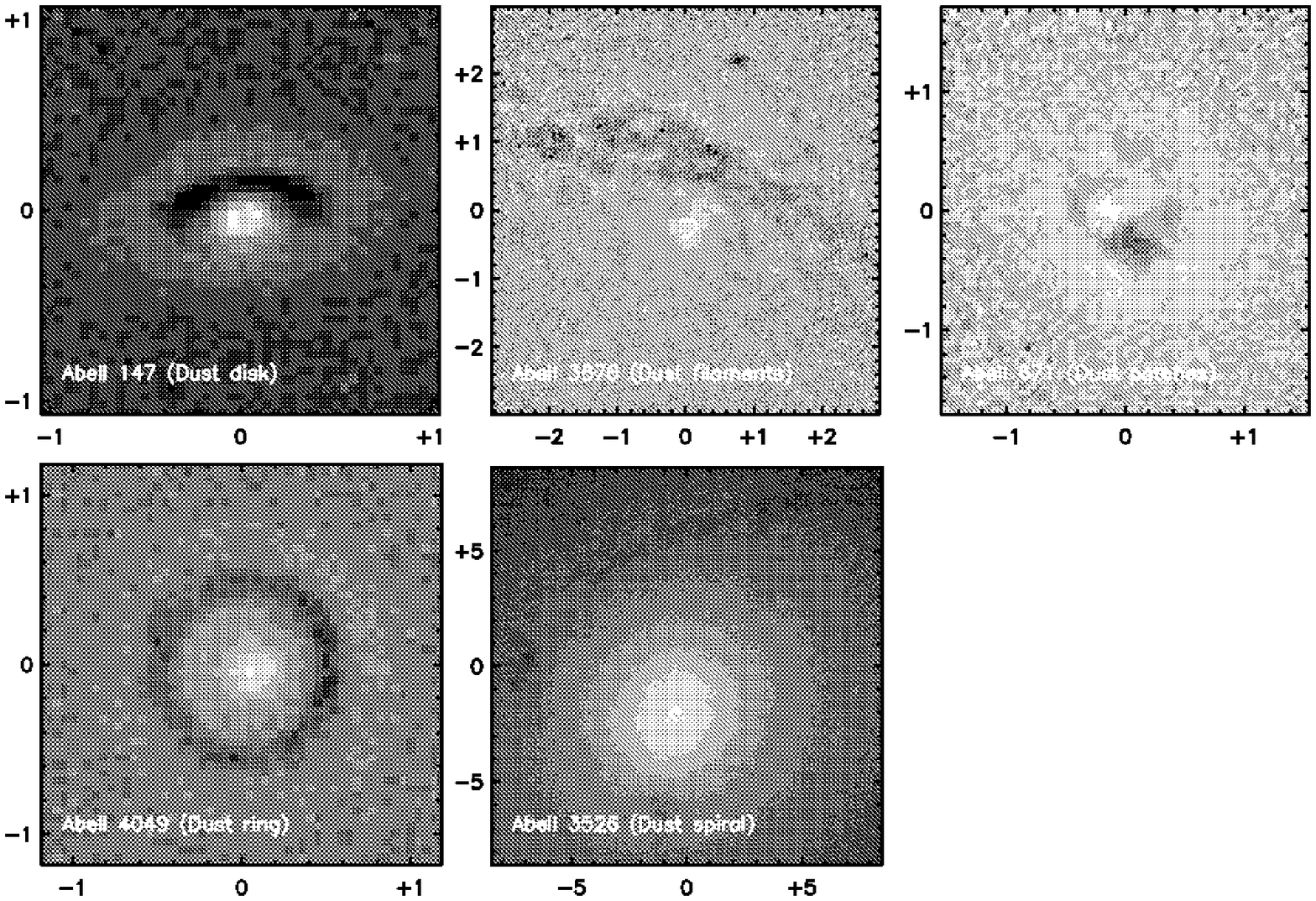}}
\ifsubmode
\addtocounter{figure}{1}
\vskip3.0truecm
\centerline{Figure~\thefigure}
\else\figcaption{\figcapdust}\fi
\end{figure}


\begin{figure}
\centerline{\psfig{figure=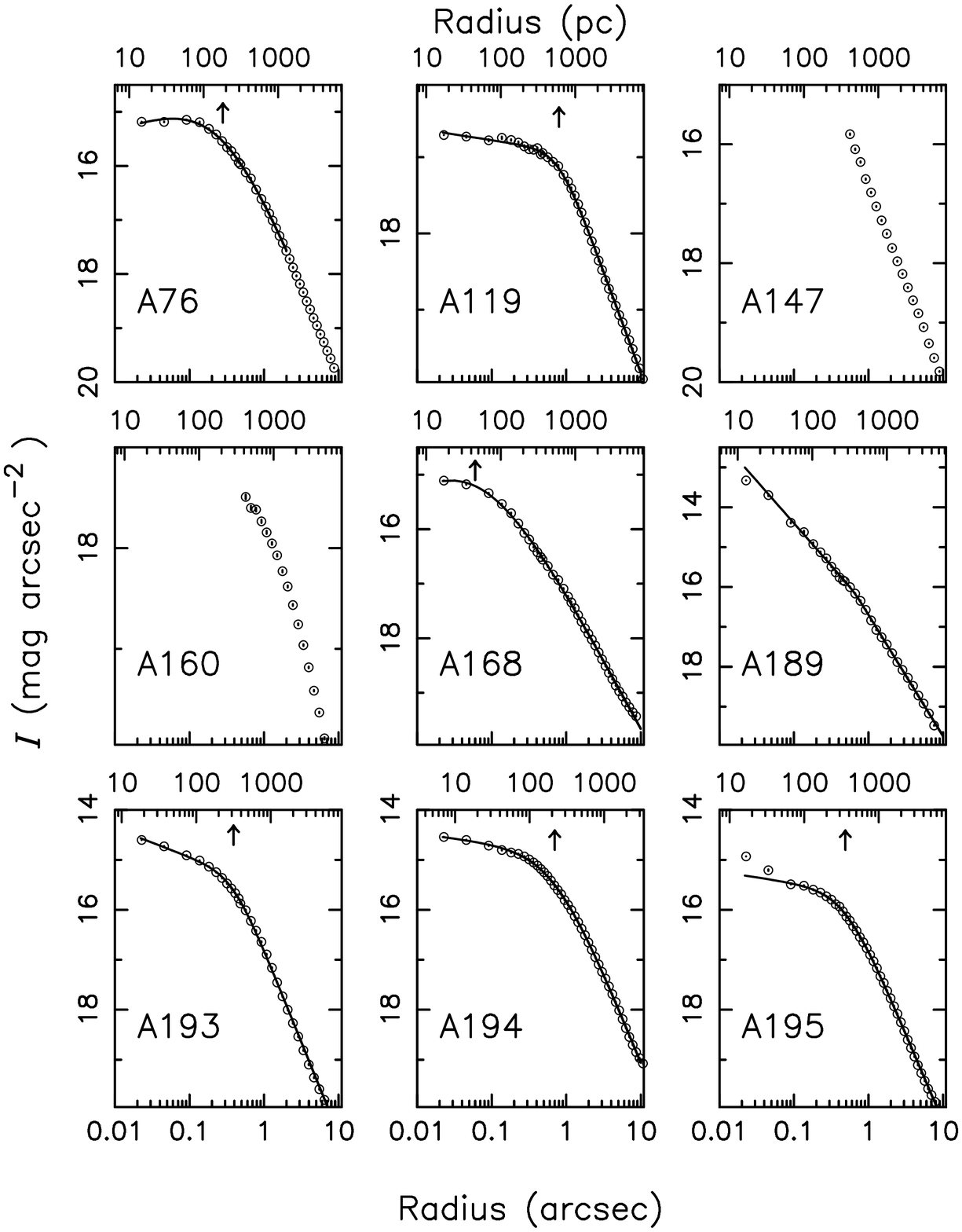,width=0.65\hsize}}
\ifsubmode
\addtocounter{figure}{1}
\vskip3.0truecm
\centerline{Figure~\thefigure}
\else\figcaption{\figcapprofiles}\fi
\end{figure}


\clearpage
\begin{figure}
\centerline{\psfig{figure=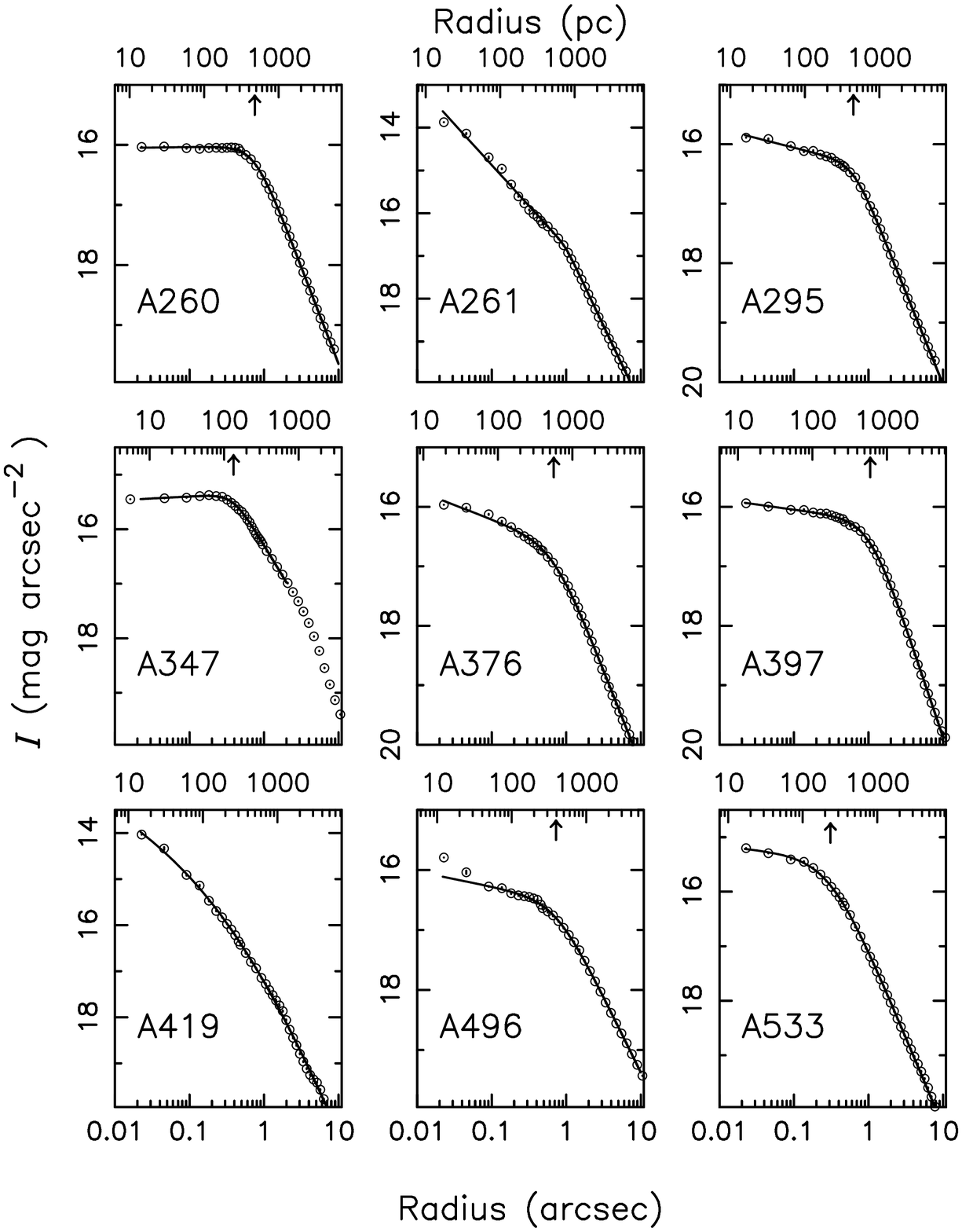,width=0.9\hsize}}
\ifsubmode
\vskip3.0truecm
\centerline{Figure~\thefigure\ (continued)}
\else
\centerline{Fig. \thefigure.--- (continued)}
\fi
\end{figure}


\clearpage
\begin{figure}
\centerline{\psfig{figure=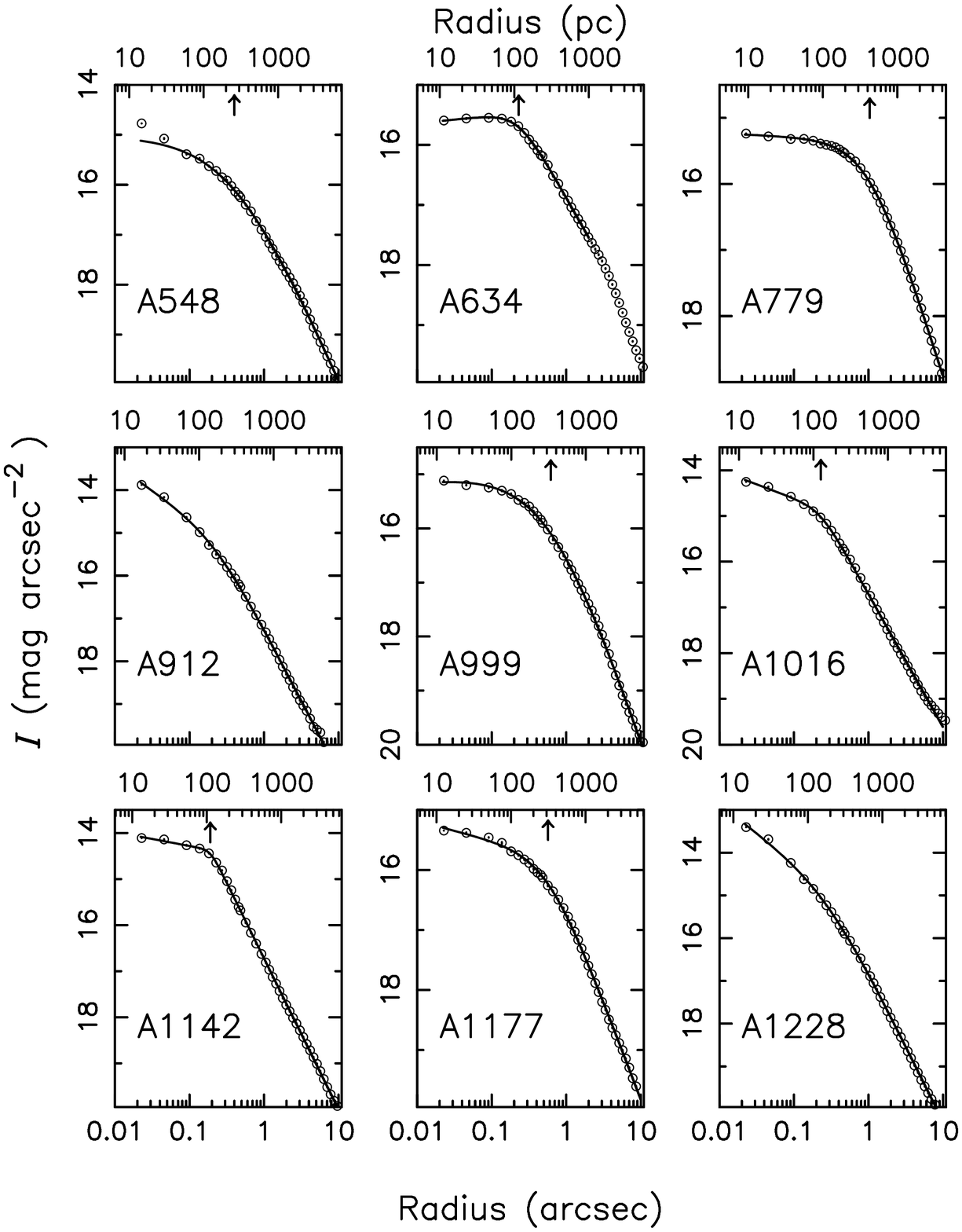,width=0.9\hsize}}
\ifsubmode
\vskip3.0truecm
\centerline{Figure~\thefigure\ (continued)}
\else
\centerline{Fig. \thefigure.--- (continued)}
\fi
\end{figure}


\clearpage
\begin{figure}
\centerline{\psfig{figure=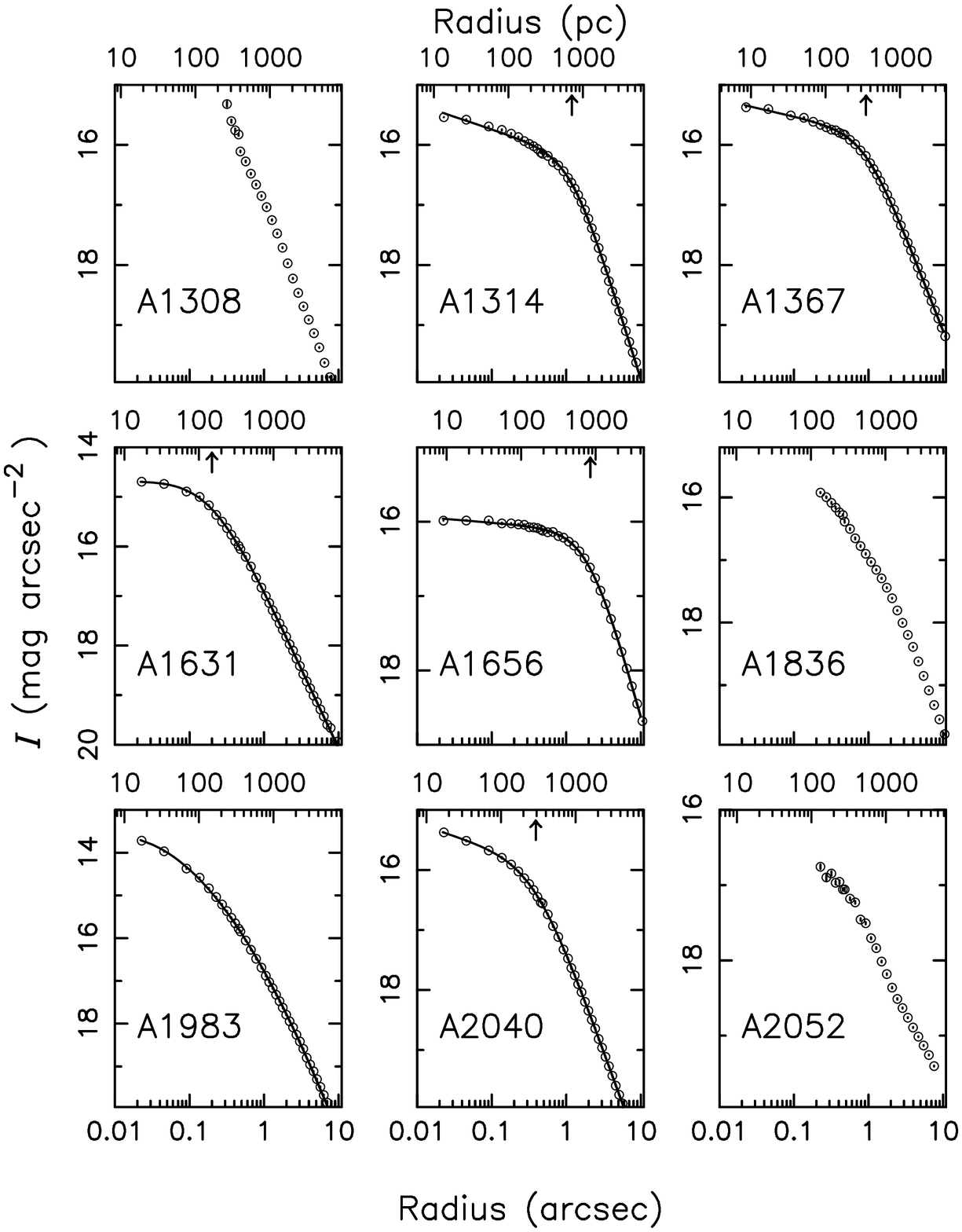,width=0.9\hsize}}
\ifsubmode
\vskip3.0truecm
\centerline{Figure~\thefigure\ (continued)}
\else
\centerline{Fig. \thefigure.--- (continued)}
\fi
\end{figure}


\clearpage
\begin{figure}
\centerline{\psfig{figure=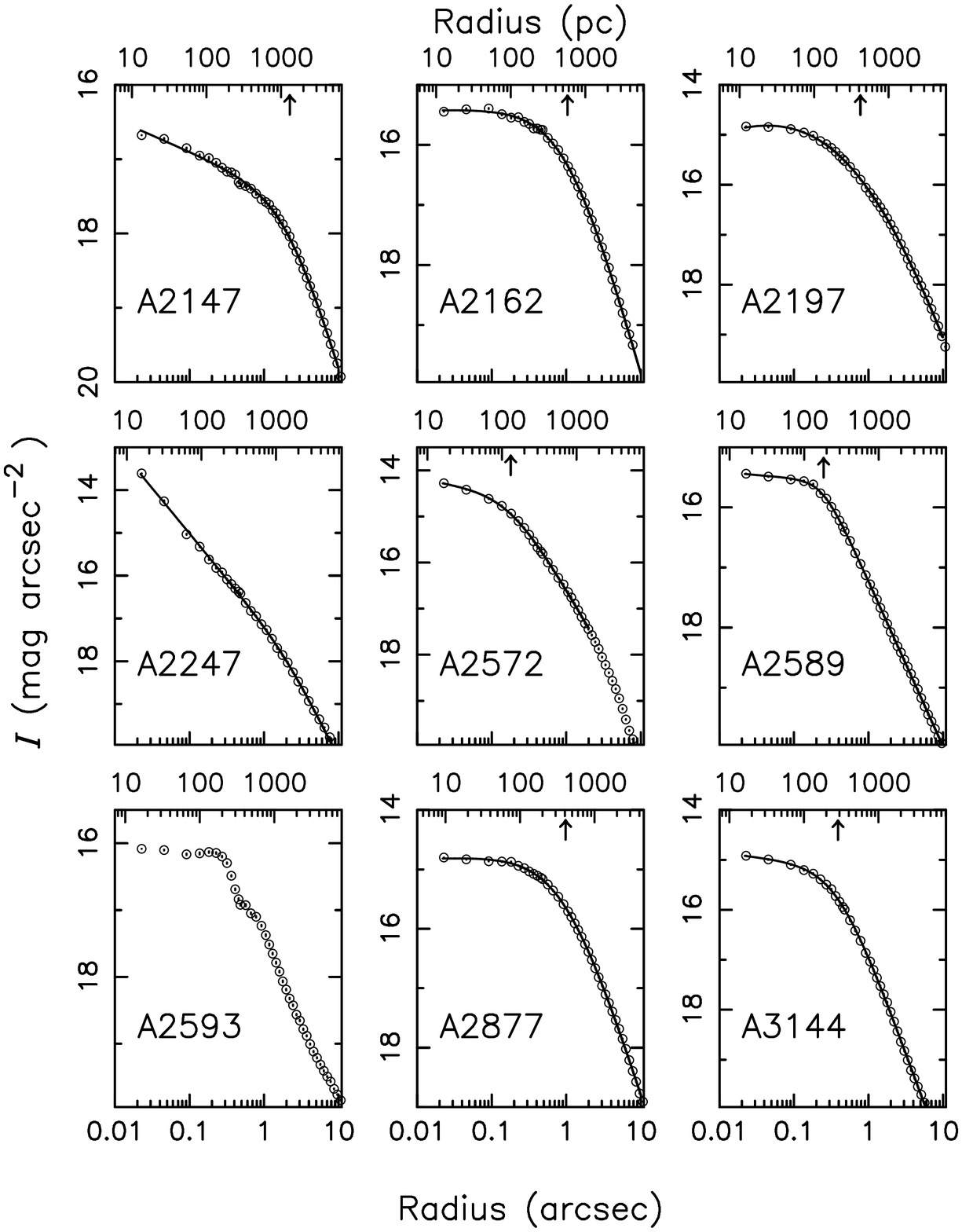,width=0.9\hsize}}
\ifsubmode
\vskip3.0truecm
\centerline{Figure~\thefigure\ (continued)}
\else
\centerline{Fig. \thefigure.--- (continued)}
\fi
\end{figure}


\clearpage
\begin{figure}
\centerline{\psfig{figure=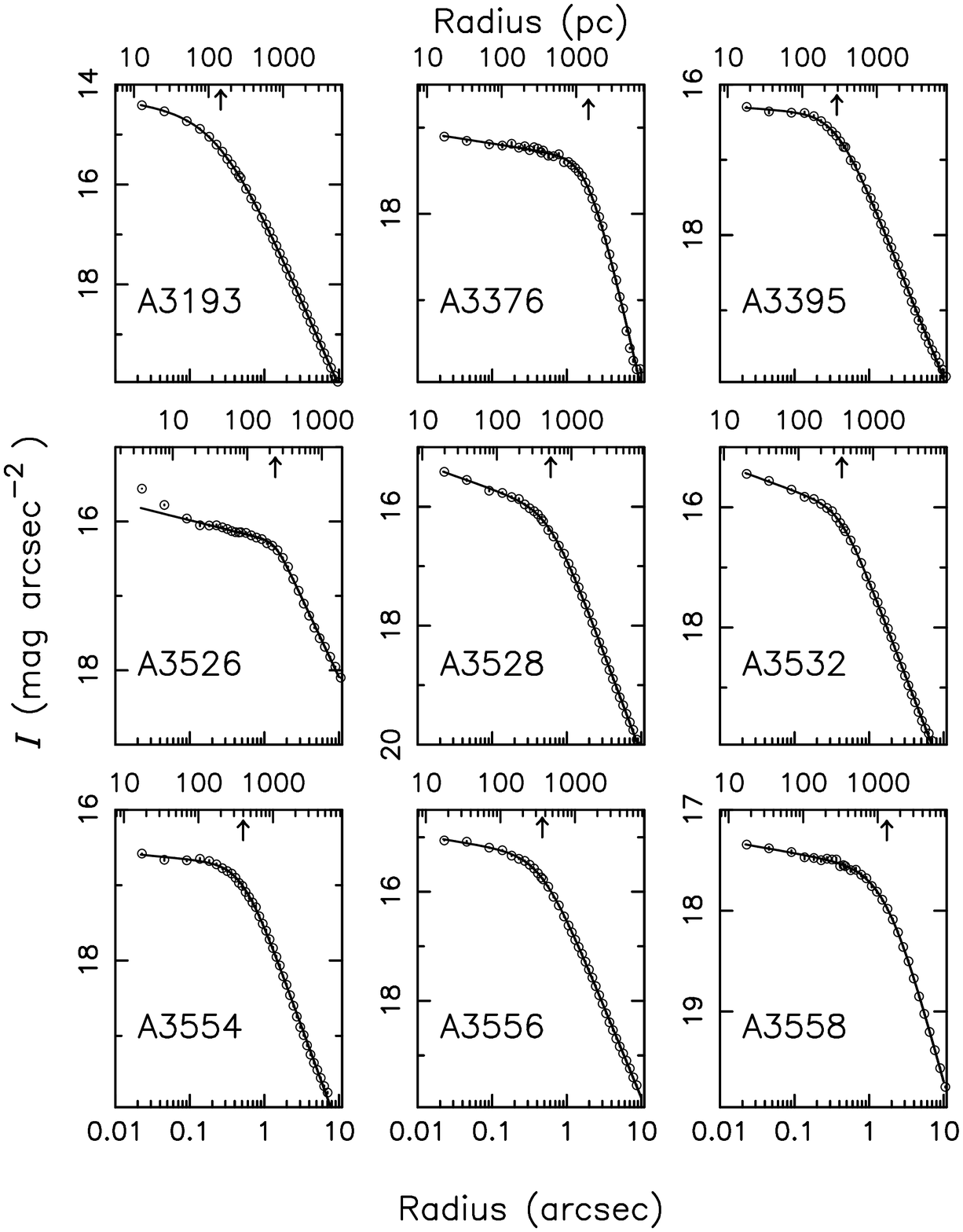,width=0.9\hsize}}
\ifsubmode
\vskip3.0truecm
\centerline{Figure~\thefigure\ (continued)}
\else
\centerline{Fig. \thefigure.--- (continued)}
\fi
\end{figure}


\clearpage
\begin{figure}
\centerline{\psfig{figure=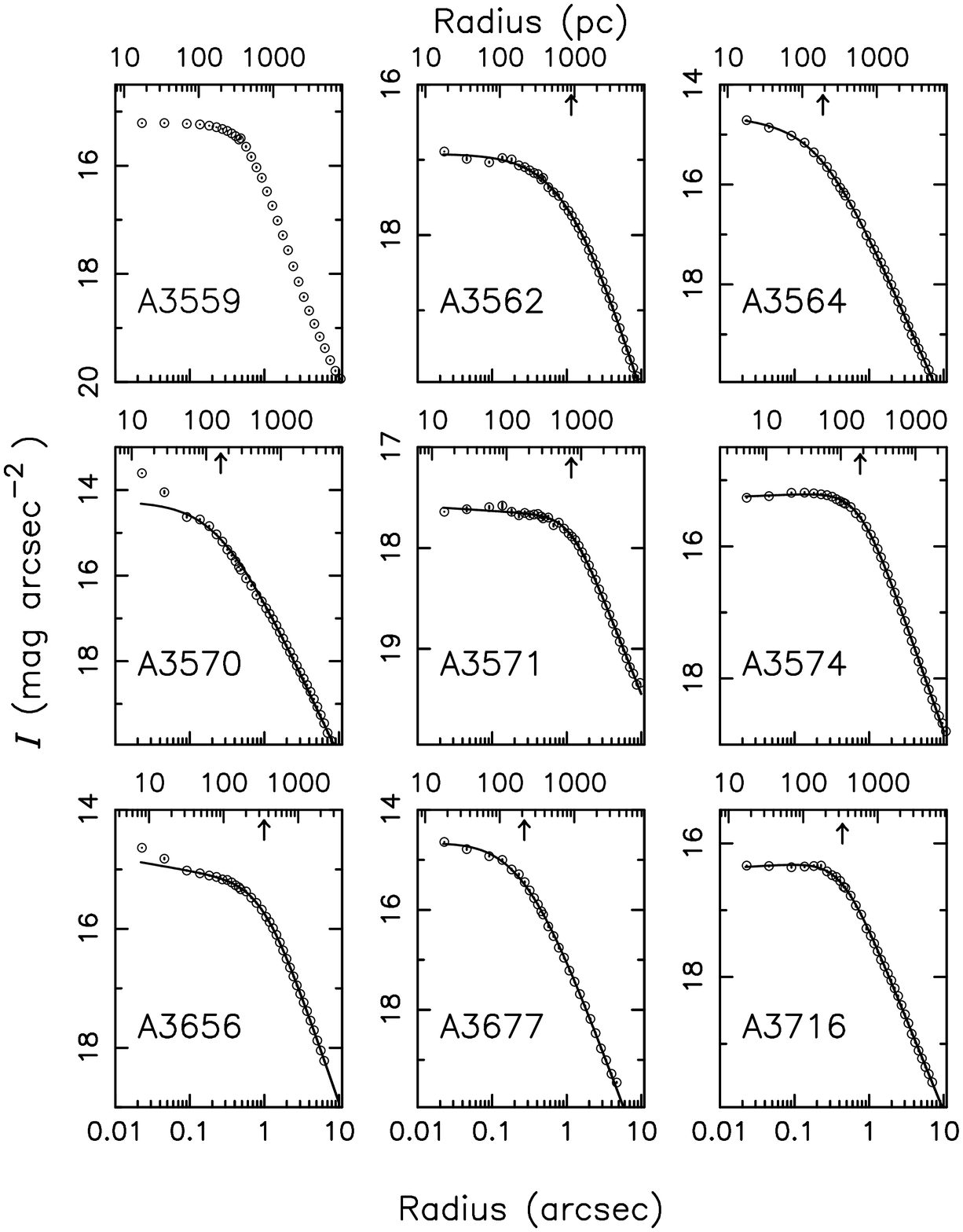,width=0.9\hsize}}
\ifsubmode
\vskip3.0truecm
\centerline{Figure~\thefigure\ (continued)}
\else
\centerline{Fig. \thefigure.--- (continued)}
\fi
\end{figure}


\clearpage
\begin{figure}
\centerline{\psfig{figure=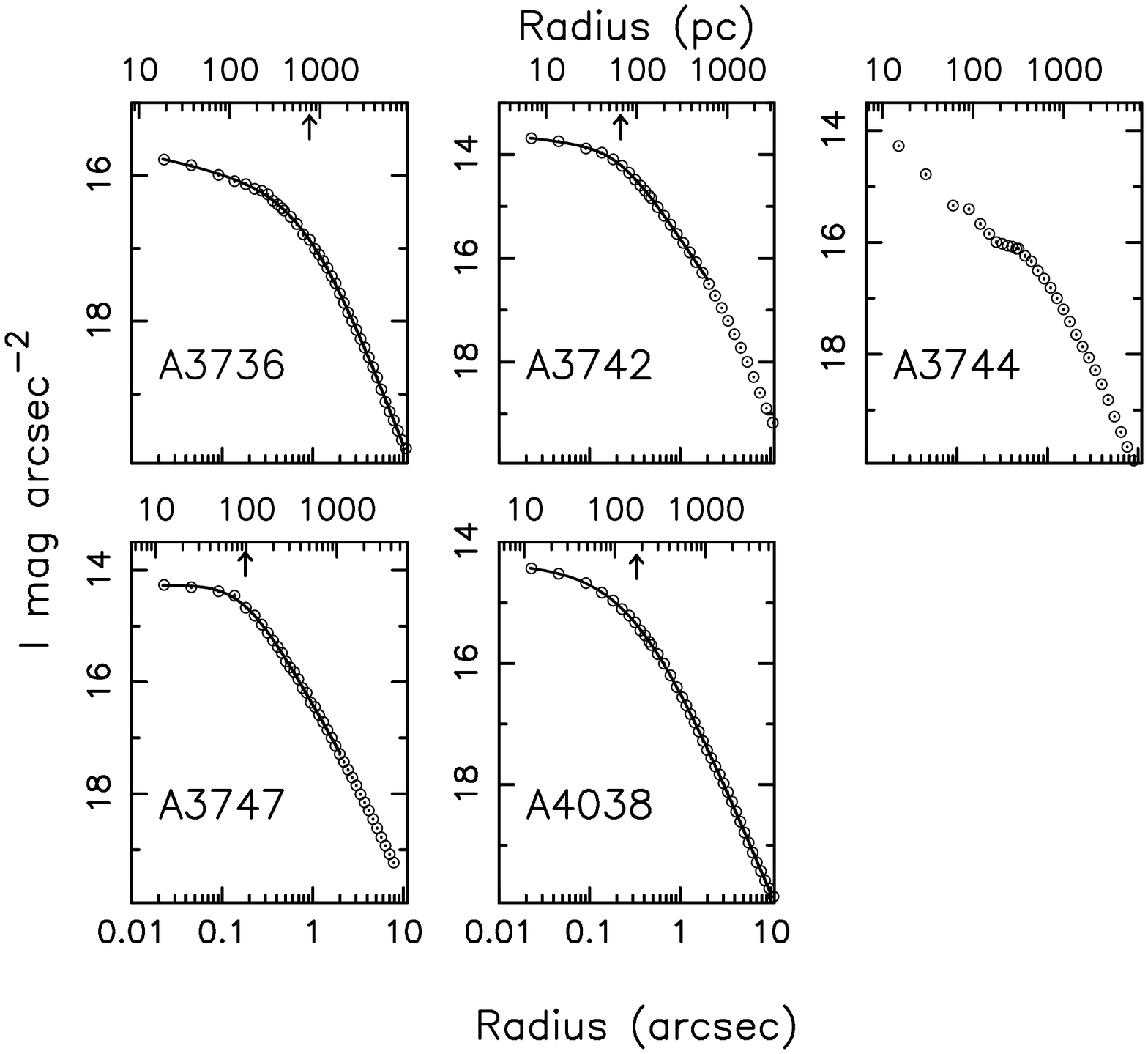,width=0.9\hsize}}
\ifsubmode
\vskip3.0truecm
\centerline{Figure~\thefigure\ (continued)}
\else
\centerline{Fig. \thefigure.--- (continued)}
\fi
\end{figure}


\begin{figure}
\epsfxsize=0.9\hsize
\centerline{\epsfbox{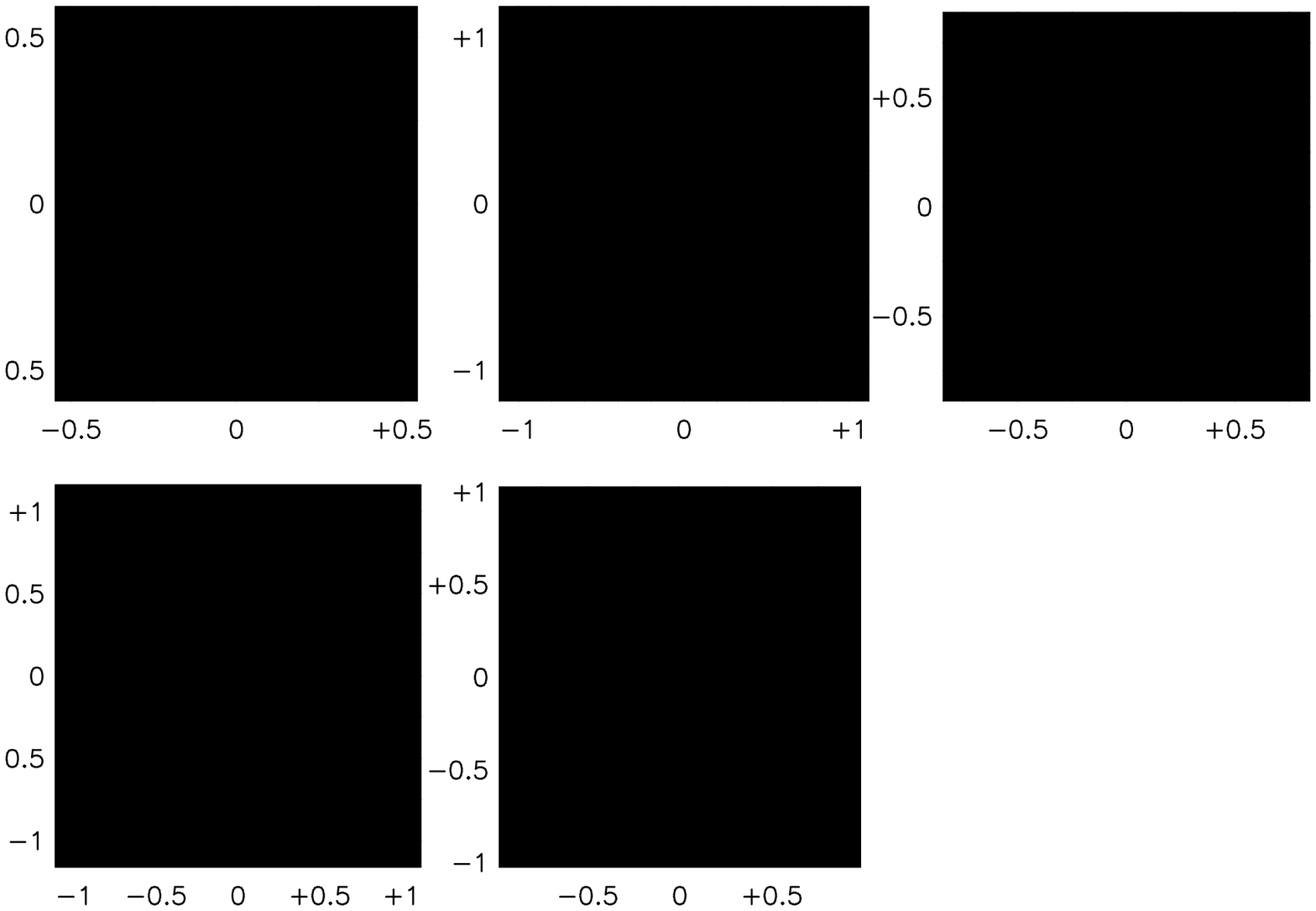}}
\ifsubmode
\addtocounter{figure}{1}
\vskip3.0truecm
\centerline{Figure~\thefigure}
\else\figcaption{\figcaphollow}\fi
\end{figure}


\begin{figure}
\centerline{\psfig{figure=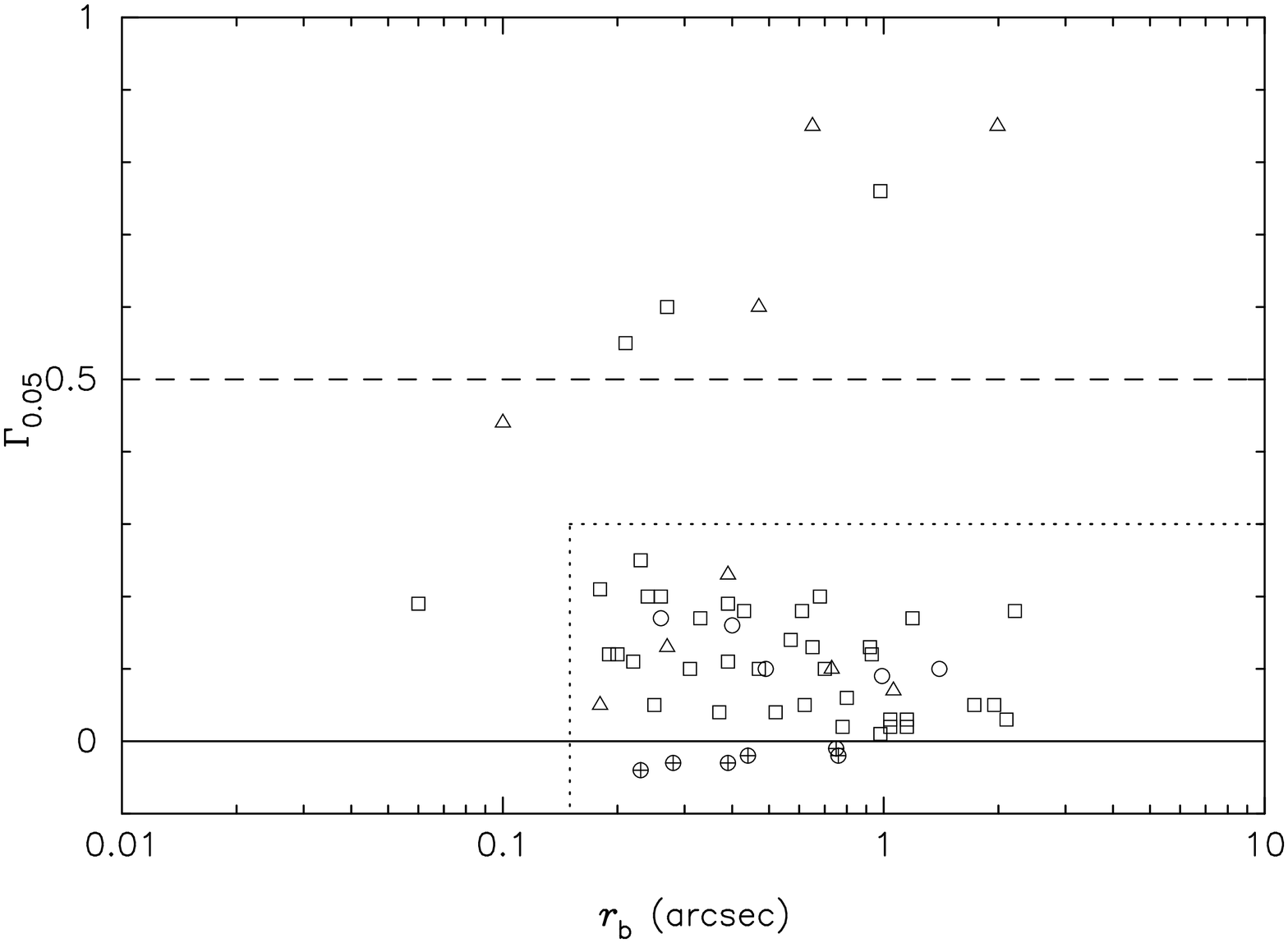,height=4.0in,angle=270}}
\ifsubmode
\addtocounter{figure}{1}
\vskip3.0truecm
\centerline{Figure~\thefigure}
\else\figcaption{\figcapgammavsrad}\fi
\end{figure}


\begin{figure}
\centerline{\psfig{figure=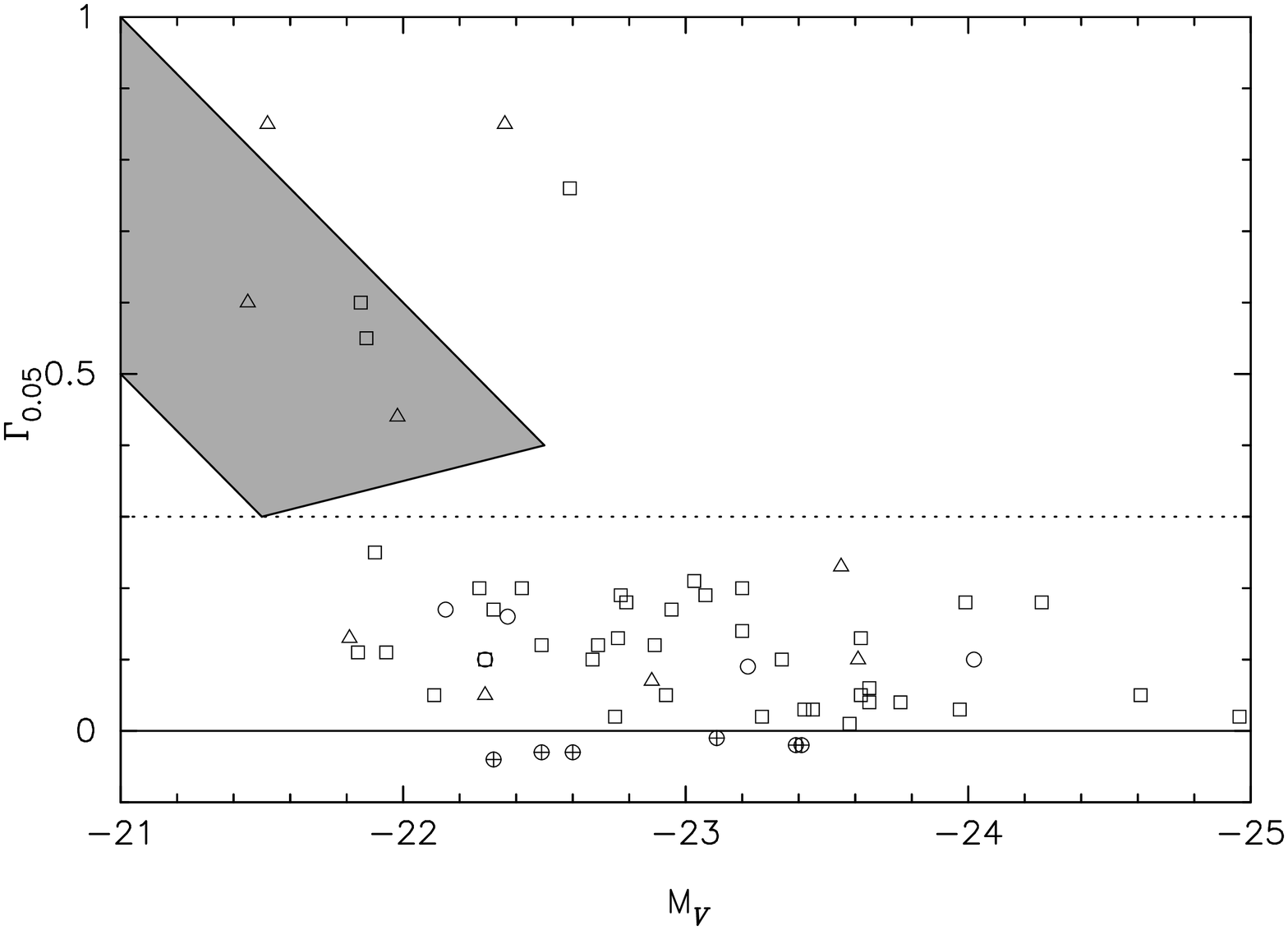,height=4.0in,angle=270}}
\ifsubmode
\addtocounter{figure}{1}
\vskip3.0truecm
\centerline{Figure~\thefigure}
\else\figcaption{\figcapgammavsMV}\fi
\end{figure}


\begin{figure}
\centerline{\psfig{figure=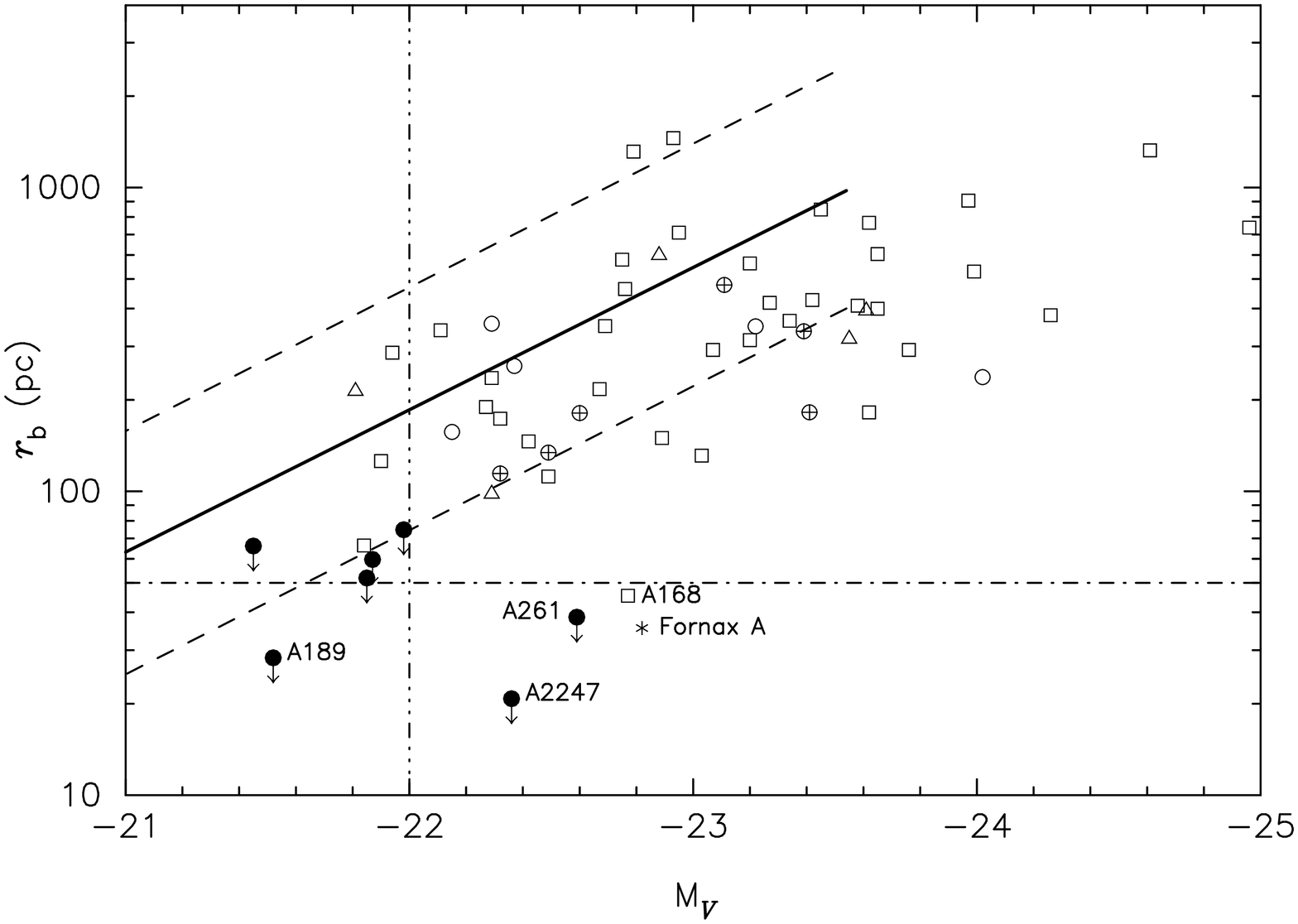,height=4.0in,angle=270}}
\ifsubmode
\addtocounter{figure}{1}
\vskip3.0truecm
\centerline{Figure~\thefigure}
\else\figcaption{\figcapradvsMV}\fi
\end{figure}


\begin{figure}
\centerline{\psfig{figure=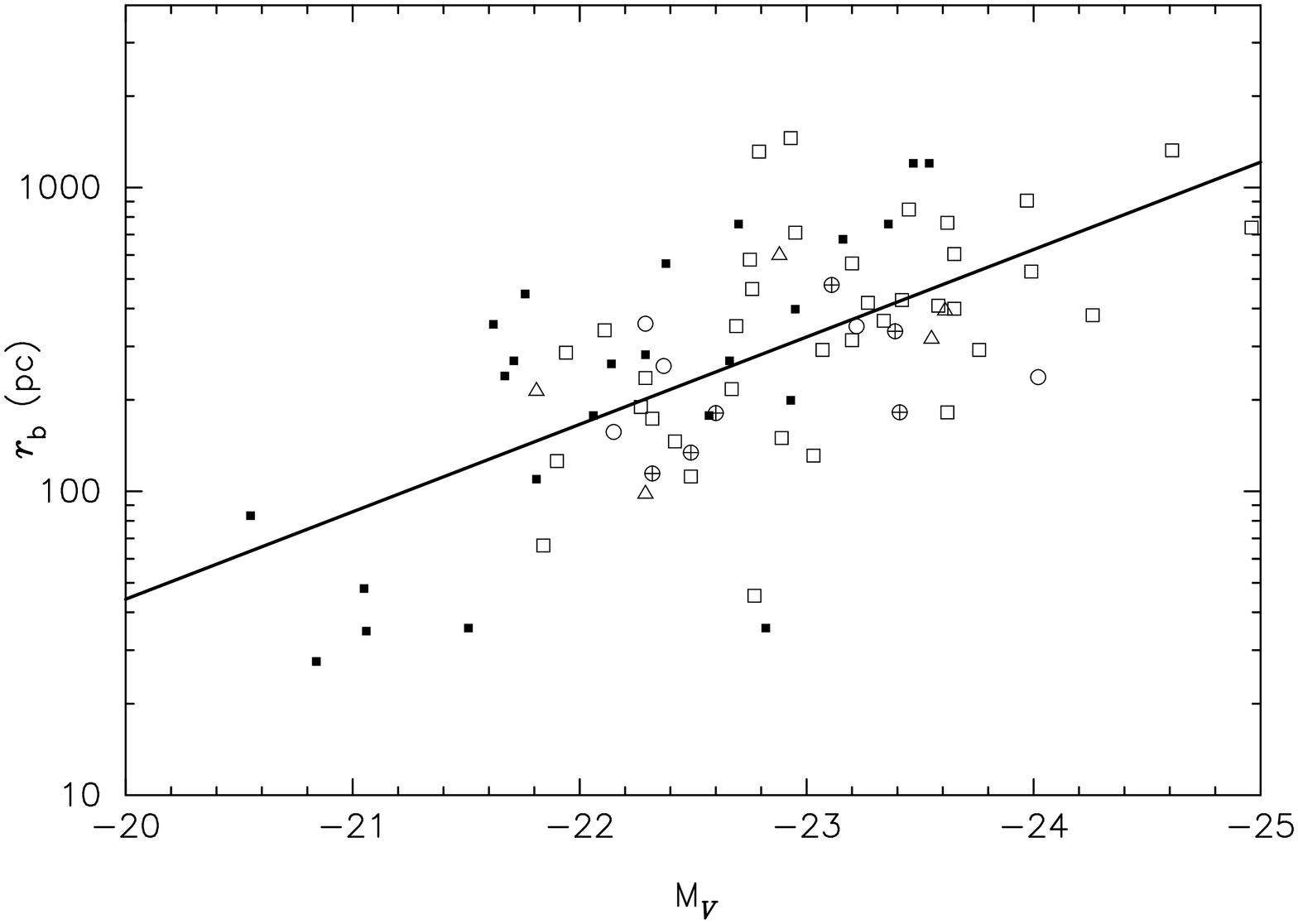,height=4.0in,angle=270}}
\ifsubmode
\addtocounter{figure}{1}
\vskip3.0truecm
\centerline{Figure~\thefigure}
\else\figcaption{\figcapradvsMVfit}\fi
\end{figure}


\begin{figure}
\centerline{\psfig{figure=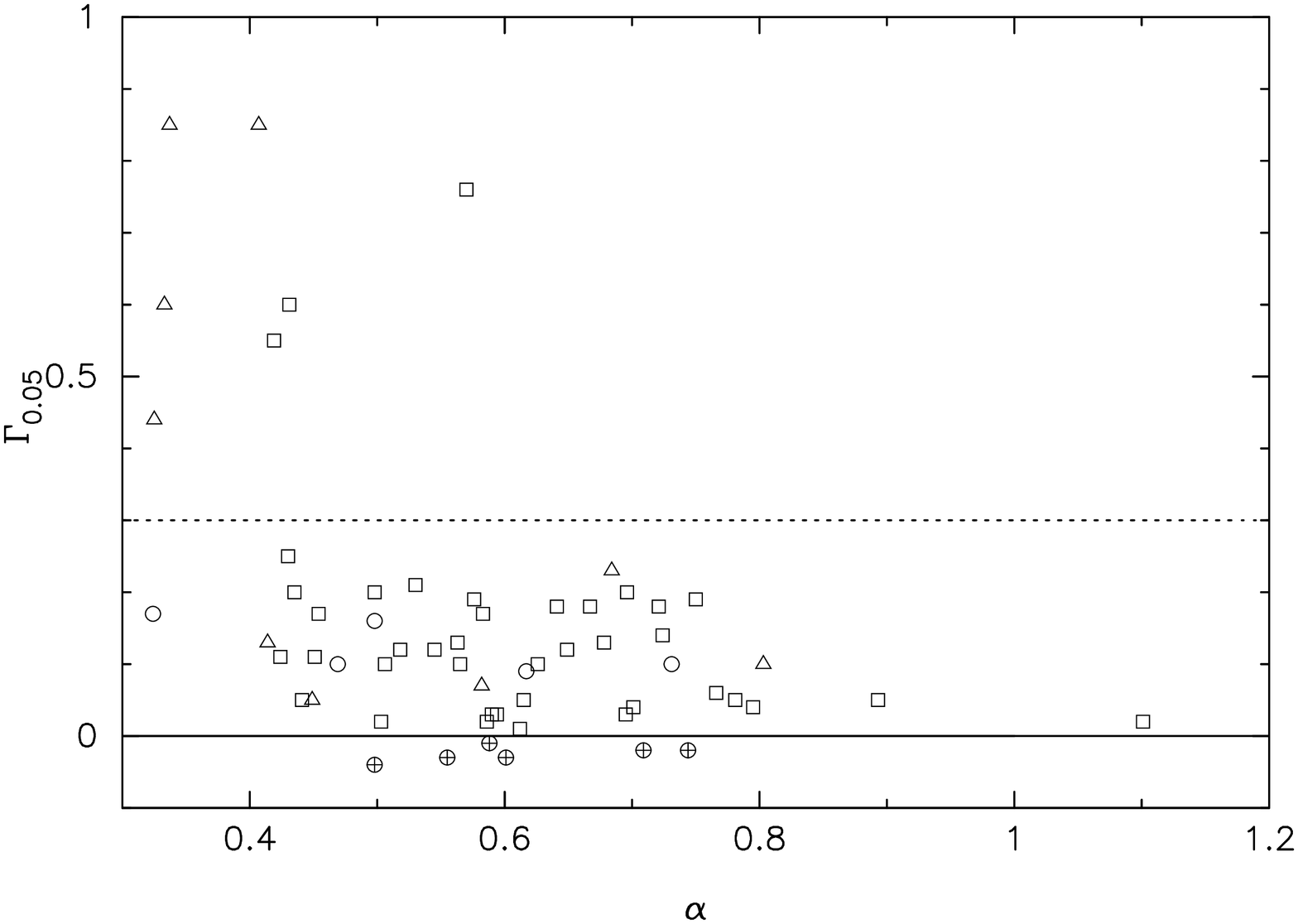,height=4.0in,angle=270}}
\ifsubmode
\addtocounter{figure}{1}
\vskip3.0truecm
\centerline{Figure~\thefigure}
\else\figcaption{\figcapgammavsalpha}\fi
\end{figure}


\fi 


\clearpage
\ifsubmode\pagestyle{empty}\fi


\begin{deluxetable}{lcccccccccccc}
\tabletypesize{\tiny}
\rotate
\tablecaption{SAMPLE OF ABELL CLUSTER BCG GALAXIES\label{t:sample}}
\tablewidth{0pt}
\tablehead{
\colhead{Cluster} & \colhead{Other name} & \colhead{R.A.} & 
\colhead{Dec.} & \colhead{$cz$} & \colhead{D} 
 & \colhead{$R_{\rm met}$} & \colhead{$M_{V}$} & \colhead{log~$L_{V}$} & \colhead{Profile} & \colhead{North} & \colhead{E($B-V$)} 
 & \colhead{Notes}\\
\colhead{}&\colhead{}&\colhead{(J2000)}&\colhead{(J2000)}& 
\colhead{(km~s$^{-1}$)} & \colhead{(Mpc)} & \colhead{}& \colhead{}& 
\colhead{($\Lsun$)} & \colhead{} & \colhead{(Figure~\ref{f:images})} & 
\colhead{} & \colhead{}    
}
\startdata 
ABELL76  & IC 1565 &      00 39 26.28  & +06 44 03.2 & 11162 & 33.37    & 13.10 & -22.60  & 10.97 & Y & -70.2 &    0.045 &Hollow\\
ABELL119 & UGC 579 &      00 56 16.11  & $-01$ 15 19.1 & 13354 &155.68   & 13.43  & -23.65  & 11.39 & Y & 107.5 &  0.039 & \nodata\\
ABELL147 & \nodata &      01 08 38.00  & +02 16 07.7 & 13022 &151.63  	  & 13.20  & -22.24  & 10.83 & S & 114.8 &   0.025 &Dust, MN\\
ABELL160 & \nodata &      01 12 59.68  & +15 29 28.7 & 13095 &153.53    & 13.96  & -22.82  & 11.06 & S & 122.8 &     0.086 &Dust, MN\\
ABELL168 & UGC 797 &      01 14 57.61  & +00 25 50.6 & 13446 &155.68    & 13.48  & -22.77  & 11.04 & Y & -73.1 &   0.028 &\nodata\\
ABELL189 & \nodata &      01 23 26.44  & +01 42 20.5 &  9925 &116.71    & 13.02  & -21.52  & 10.54 & Y & -63.0 &   0.033 &Dust, MN\\
ABELL193 & IC 1695 &      01 25 07.62  & +08 41 58.4 & 14553 &167.98    & 13.69  & -23.55  & 11.35 & Y  & -66.9 &  0.051 &Dust, MN\\
ABELL194 & NGC 545 &      01 25 59.24  & $-01$ 20 25.4 &  5357 & 64.10    & 11.67  & -22.67  & 11.00 & Y  & -69.0 &0.040 &\nodata\\
ABELL195 &  IC 115 &      01 26 54.46  & +19 12 52.3 & 12796 &150.16     & 13.35 & -22.29  & 10.85 & Y  & -63.7 &  0.056 &Nuc\\
ABELL260 & IC 1733 &      01 50 42.91  & +33 04 55.5 & 10998 &131.58    & 12.92  & -23.11  & 11.18 & Y  & 98.6 &   0.046 &Hollow\\
ABELL261 & \nodata &      01 51 27.23  & $-02$ 15 32.1 & 13965 &158.89    & 13.56  & -22.59  & 10.97 & Y  & 39.7 & 0.029 &Disk, MN\\
ABELL262 & NGC 708 &      01 52 46.41  & +36 09 08.2 &  4913 & 62.63    & 11.90  & -22.45  & 10.91 & N & 107.4 &     0.086 &Dust\\
ABELL295 & UGC 1525 &     02 02 17.25  & $-01$ 07 40.0 & 12890 &147.10    & 13.41 & -22.76  & 11.04 & Y  & -70.2 & 0.027 &\nodata\\
ABELL347 & NGC 910 &      02 25 26.79  & +41 49 26.4 &  5604 & 70.95    & 11.89  & -22.49  & 10.93 & Y & -175.1 &    0.058 &Double, Hollow\\
ABELL376 & UGC 2232 &     02 46 04.11  & +36 54 19.6 & 14700 &170.75    & 13.81 & -23.20  & 11.21 & Y  & 128.4 &   0.079 &\nodata\\
ABELL397 & UGC 2413 &     02 56 28.79  & +15 54 57.8 &  9975 &116.34   & 12.90 & -22.88  & 11.08 & Y  &  -59.0 &   0.141 &Dust\\
ABELL419 & \nodata &      03 08 28.83  & $-23$ 36 53.0 & 12268 &136.49    & 13.55  & -21.45  & 10.51 & Y  & -95.3 &0.023 &Dust, MN\\
ABELL496 & \nodata &      04 33 37.75  & $-13$ 15 42.4 &  9893 &110.97    & 12.89  & -23.61  & 11.38 & Y  & 123.4 &0.136 &Dust, MN, Nuc\\
ABELL533 & \nodata &      05 01 08.30  & $-22$ 34 58.4 & 14149 &156.97   & 13.59  & -22.29  & 10.85 & Y  & -53.0 & 0.035 &MN\\
ABELL548 & ESO 488--G33 & 05 49 21.64  & $-25$ 20 47.7 & 11949&133.52  & 13.13  & -22.37  & 10.88 & Y  & 87.4 &    0.029 &Nuc\\
ABELL569 & NGC 2329 &     07 09 08.01  & +48 36 55.5 &  5749 & 73.32    & 11.65 & -22.56  & 10.96 & N & -45.4 &     0.071 &Dust, Nuc\\
ABELL634 & UGC 4289 &     08 15 44.75  & +58 19 15.6 &  8135 &102.62    & 12.56 & -22.32  & 10.86 & Y  & -171.5 &  0.059 &Hollow\\
ABELL671 & IC 2378 &      08 28 31.65  & +30 25 52.4 & 15219 &176.18    & 13.45  & -23.67  & 11.40 & N & 166.1 &     0.047 &Dust, MN\\
ABELL779 & NGC 2832 &     09 19 46.83  & +33 44 58.8 &  6796 & 84.66    & 11.56 & -23.42  & 11.30 & Y  & 175.6 &   0.017 &MN\\
ABELL912 & \nodata &      10 01 09.47  & $-00$ 04 46.8 & 13572 &153.96    & 13.80  & -21.87  & 10.68 & Y  & 104.9 &0.037 &MN\\
ABELL999 & \nodata &      10 23 23.82  & +12 50 05.7 &  9603 &112.82    & 12.77  & -22.11  & 10.78 & Y  & -24.7 &  0.041 &MN\\
ABELL1016 & IC 613 &      10 27 07.81  & +11 00 38.3 &  9669 &112.81    & 12.97  & -21.90  & 10.69 & Y  & -22.6 &  0.031 &\nodata\\
ABELL1060 & NGC 3311 &    10 36 43.08  & $-27$ 31 36.0 & 3719 & 38.63  	  & 11.01 & -23.41  & 11.30 & N & -20.1 &   0.079 &Dust\\
ABELL1142 & IC 664 &      11 00 45.34  & +10 33 10.9 & 10501 &121.48    & 13.09  & -22.49  & 10.93 & Y  & 162.9 &  0.029 &Disk, MN\\
ABELL1177 & NGC 3551 &    11 09 44.41  & +21 45 31.8 &  9551 &113.81    & 12.90& -23.20  & 11.21 & Y  & 170.4 &    0.018 &\nodata\\
ABELL1228 & IC 2738 &     11 21 23.05  & +34 21 24.1 & 10973 &133.56    & 13.16 & -21.85  & 10.67 & Y  & 2.3 &     0.024 &MN\\
ABELL1308 & \nodata &     11 33 05.06  & $-04$ 00 48.5 & 15508 &172.42    & 13.49 & -23.08  & 11.16 & S & 8.2 &     0.046 &Dust\\
ABELL1314 & IC 712 &      11 34 49.31  & +49 04 39.8 &  9838 &123.14   & 12.84  & -22.95  & 11.11 & Y  & 147.7 &   0.016 &MN\\
ABELL1367 & NGC 3842 &    11 44 02.13  & +19 56 59.7 &  6469 & 77.58   & 11.77& -22.69  & 11.01 & Y  & -77.7 &     0.021 &\nodata\\
ABELL1631 & \nodata &     12 53 18.37  & $-15$ 32 04.4 & 14030 &154.63    & 13.56 & -22.89  & 11.09 & Y  & 26.6 &  0.054 &MN\\
ABELL1656 & NGC 4889 &    13 00 08.12  & +27 58 36.7 &  6961 & 83.06    & 11.55& -23.45  & 11.31 & Y  & 158.4 &    0.010 &\nodata\\
ABELL1836 & \nodata &     14 01 41.85  & $-11$ 36 25.0 & 11036 &122.33    & 12.87 & -22.98  & 11.12 & S & -20.3 &    0.064 &Dust\\
ABELL1983 & \nodata &     14 52 43.25  & +16 54 13.5 & 13617 &154.07    & 13.27 & -21.98  & 10.72 & Y  & 155.4 &   0.027 &Dust, MN\\
ABELL2040 & UGC 9767 &    15 12 47.51  & +07 26 03.0 & 13698 &154.46    & 14.04& -23.07  & 11.16 & Y  & -82.0 &    0.042 &\nodata\\
ABELL2052 & UGC 9799 &    15 16 44.59  & +07 01 17.6 & 10575 &120.21    & 13.05& -24.07  & 11.56 & S & -81.6 &       0.037 &Dust, Nuc, MN\\
ABELL2147 & UGC 10143 &   16 02 17.03  & +15 58 28.8 & 10511 &122.47   & 13.20& -22.79  & 11.05 & Y  & 173.5 &     0.031 &\nodata\\
ABELL2162 & NGC 6086 &    16 12 35.55  & +29 29 05.4 &  9629 &114.82    & 12.58& -22.75  & 11.03 & Y  & -118.8 &   0.037 &\nodata\\
ABELL2197 & NGC 6173 &    16 29 44.87  & +40 48 41.7 &  9042 &110.41    & 12.18& -23.27  & 11.24 & Y  & 148.2 &    0.007 &\nodata\\
ABELL2247 & UGC 10638 &   16 52 47.76  & +81 37 58.1 & 11641 &142.86 & 13.13  & -22.36  & 10.88 & Y  & -178.4 &    0.063 &Dust\\
ABELL2572 & NGC 7578B &   23 17 13.55  & +18 42 29.1 & 12420 &150.42   & 13.20& -23.03  & 11.14 & Y  & -153.3 &    0.067 &\nodata\\
ABELL2589 & NGC 7647 &    23 23 57.40  & +16 46 38.2 & 12397 &150.03    & 13.59& -23.62  & 11.38 & Y  & -72.8 &    0.030 &\nodata\\
ABELL2593 & NGC 7649 &    23 24 20.10  & +14 38 49.2 & 12489 &150.75    & 13.52& -23.39  & 11.29 & S & 117.2 &       0.044 &Dust, MN\\
ABELL2634 & NGC 7720 &    23 38 29.40  & +27 01 53.2 &  9153 &113.97    & 12.53& -23.46  & 11.32 & N & -73.8 &       0.070 &Dust, Nuc, MN\\
ABELL2657 & \nodata &     23 44 30.44  & +09 15 50.2 & 12100 &146.25    & 13.39 & -22.03  & 10.74 & N & 114.7 &      0.126 &Dust, MN\\
ABELL2666 & NGC 7768 &    23 50 58.56  & +27 08 50.0 &  8057 &101.23    & 12.13& -22.87  & 11.08 & N & 99.4 &        0.039 &Dust\\
ABELL2877 & IC 1633 &     01 09 55.53  & $-45$ 55 53.0 &  7309 & 86.10    & 11.48 & -23.58  & 11.36 & Y  & -123.7 &0.011 &\nodata\\
ABELL3144 & \nodata &     03 37 05.66  & $-55$ 01 19.0 & 13546 &151.41    & 13.68 & -21.94  & 10.71 & Y  & 152.1 & 0.016 &MN\\
ABELL3193 & NGC 1500 &    03 58 13.93  & $-52$ 19 42.2 & 10350 &115.88    & 12.90& -22.42  & 10.90 & Y  & 122.9 &  0.013 &\nodata\\
ABELL3376 & ESO 307--G13 &06 00 41.05  & $-40$ 02 41.1 & 13907 &154.03 & 13.46 & -22.93  & 11.10 & Y  & 146.7 &    0.052 &\nodata\\
ABELL3395 & ESO 161--G08 &06 27 36.23  & $-54$ 26 58.2 & 14712 &162.85 &13.88 & -23.76  & 11.44 & Y  & 95.4 &      0.113 &\nodata\\
ABELL3526 & NGC 4696 &    12 48 49.18  & $-41$ 18 40.5 & 3454 & 35.04   & 10.12  & -24.02  & 11.54 & Y & 156.4 &     0.114 &Double, Dust, Nuc\\
ABELL3528 & ESO 443--G04 &12 54 22.28  & $-29$ 00 47.0 & 16317 &178.98&13.55& -23.99  & 11.53 & Y  & 32.9 & 0.077 & MN\\
ABELL3532 & \nodata &     12 57 21.96  & $-30$ 21 48.7 & 16646 &182.47    & 13.89 & -24.26  & 11.64 & Y  & 71.8 &  0.085 &MN\\
ABELL3554 & ESO 382--G43 &13 19 31.53  & $-33$ 29 17.2 & 14333 &158.29& 13.82 & -23.65  & 11.39 & Y  & 22.9 &      0.064 &MN\\
ABELL3556 & ESO 444--G25 &13 24 06.69  & $-31$ 40 12.7 & 14500 &160.12& 13.15& -23.34  & 11.27 & Y  & 16.9 &       0.059 &MN\\
ABELL3558 & ESO 444--G46 &13 27 56.80  & $-31$ 29 45.7 & 14312 &158.20& 13.18& -24.61  & 11.78 & Y  & -4.6 &       0.050 &MN\\
ABELL3559 & ESO 444--G55 & 13 29 51.31 & $-29$ 30 48.8 & 14213 &157.12 & 13.22& -23.50 & 11.33 & S & 15.3  &      0.056 &Dust\\
ABELL3562 & ESO 444--G72 &13 33 34.68  & $-31$ 40 20.6 & 14708 &162.37& 13.70& -23.97  & 11.52 & Y  & 36.1 &       0.058 &MN\\
ABELL3564 & \nodata &     13 34 55.29  & $-35$ 05 59.1 & 14721 &162.90    & 13.63 & -22.27  & 10.84 & Y  & -31.7 & 0.062 &MN\\
ABELL3565 & IC 4296 &     13 36 39.05  & $-33$ 57 58.0 &  3834 & 39.72    & 10.46 & -23.03  & 11.14 & N & -24.6 &    0.062 &Dust\\
ABELL3570 & ESO 325--G16 &13 46 23.99  & $-37$ 58 16.5 & 11156 &124.45& 12.87 & -22.15  & 10.79 & Y  & -8.6 &      0.078 &Nuc\\
ABELL3571 & ESO 383--G76 &13 47 28.43  & $-32$ 51 53.5 & 11913 &132.39& 12.74& -24.96  & 11.92 & Y  & 8.6 & 	   0.054 &\nodata\\
ABELL3574 & IC 4329 &     13 49 05.25  & $-30$ 17 45.2 &  4657 & 49.59   & 11.19 & -23.41  & 11.30 & Y  & 42.4 &   0.061 &Hollow\\
ABELL3656 & IC 4931 &     20 00 50.38  & $-38$ 34 29.8 &  5768 & 72.90    & 11.44 & -23.22  & 11.22 & Y  & 14.3 &  0.071 &Nuc\\
ABELL3676 & ESO 340--G25 &20 24 24.55  & $-40$ 21 59.1 & 12108 &144.67& 13.09& -22.70  & 11.01 & N & -70.9 &         0.043 &Dust, Sp?\\
ABELL3677 & \nodata &     20 26 23.83  & $-33$ 21 03.7 & 13789 &163.35    & 13.84 & -21.81  & 10.66 & Y  & 0.1 &   0.070 &Dust\\
ABELL3698 & NGC 6936 &    20 35 56.30  & $-25$ 16 48.0 &  6040 & 78.28    & 11.75& -21.93  & 10.70 & N & -65.0 &     0.045 &Dust\\
ABELL3716 & ESO 187--G26 &20 51 56.89  & $-52$ 37 48.8 & 13426 &157.65& 13.55& -23.39  & 11.29 & Y  & -93.8 &      0.033 &Hollow\\
ABELL3733 & NGC 6999 &    21 01 59.56  & $-28$ 03 32.3 & 11039 &134.79   & 13.49& -22.67  & 11.00 & N & -56.3 &   0.115 &Dust\\
ABELL3736 & ESO 286--G41 &21 05 04.45  & $-43$ 25 12.0 & 14604 &171.65& 13.31& -23.62  & 11.38 & Y  & 18.3 &       0.031 &MN\\
ABELL3742 & NGC 7014 &    21 07 52.18  & $-47$ 10 44.2 &  4842 & 62.19    & 11.48& -21.84  & 10.67 & Y  & -65.5 &  0.033 &MN\\
ABELL3744 & NGC 7016 &    21 07 16.32  & $-25$ 28 08.6 & 11153 &136.58    & 12.80& -22.50  & 10.93 & S & 110.9 &     0.064 &Dust, MN, Nuc?\\
ABELL3747 & ESO 286--G59 &21 08 39.11  & $-43$ 29 10.2 &  9170 &112.47& 12.67& -22.29  & 10.85 & Y  & -65.6 &      0.031 &Dust\\
ABELL4038 & IC 5353 &     23 47 28.64  & $-28$ 06 33.3 &  8501 &108.51    & 12.46 & -22.32  & 10.86 & Y  & -73.1 & 0.019 &\nodata\\
ABELL4049 & IC 5362 &     23 51 36.77  & $-28$ 21 53.9 &  8512 &108.85    & 12.43 & -22.65  & 10.99 & N & -55.9 &    0.020 &Dust\\
ABELL4059 & ESO 349--G10 &23 57 00.56  & $-34$ 45 33.8 & 14730&174.84 &13.44 & -24.08  & 11.56 & N & -103.3 &        0.017 &Dust\\
\enddata
\tablecomments{\tiny
Column~(1) lists the identification of the cluster in the catalogs of
\citet{abell58} and \citet{abell89}. Column~(2) provides other names for the
observed BCG, if any. Columns~(3) and~(4) list the coordinates of the BCG
center, accurate to 1--2 arcsec. These were generally taken from the Guide
Star Catalog, which identifies and lists most of the sample galaxies as
extended sources. For a few galaxies there is no identification in the Guide
Star Catalog, or the Guide Star Catalog coordinates are obviously in error
(e.g., due to the presence of multiple nuclei). In these cases the coordinates
were determined by applying a centroiding algorithm to Digitized Sky Survey
images. The units of the coordinates are hours, minutes, and seconds for Right
Ascension (Column 3) and degrees, arcminutes, and arcseconds for Declination
(Column 4). Column~(5) lists the heliocentric velocity $cz$ for the BCG. We
used either the measured BCG value or the average value for the cluster (in
cases where we suspect that the BCG has a substantial random velocity with
respect to the cluster center), taken from \citet{pos95} and converted it into
a CMB frame redshift, to translate arcseconds into parsecs. Column~(6) lists
the corresponding angular diameter distance, calculated using the cosmology
described in the text. Column~(7) lists the $R$-band metric magnitude in an
aperture of 10~$h^{-1}$ kpc \citep[from][]{pos95}. Column~(8) lists the total
absolute $V$-band magnitude, calculated as described in the text. Column~(9)
lists the corresponding luminosity in solar $V$-band units. Column~(10) has a
``Y'' for those galaxies for which an accurate surface brightness profile
could be determined and fitted with the Nuker formulation. This column has an
``S'' if we were able to derive a surface brightness profile which was not
accurate enough for fitting, and an ``N'' if it was impossible to derive any
reasonable surface brightness profile. The profiles are shown in
Figure~\ref{f:profiles}. Column~(11) defines the orientation on the sky of the
images that are shown in Figure~\ref{f:images}. The listed value is the
direction of north, measured in degrees clockwise from the axis that points
towards the top of the page. Column~(12) lists the foreground reddening
E($B-V$),  based on the work of \citet{schl98}. Column~(13) identifies any
special features seen in the {\it HST} images, using the following
abbreviations:  `Sp': spiral galaxy morphology; `MN': Multiple nuclei (see
Section~\ref{ss:multiple}); `Double': double nucleus (see
Section~\ref{ss:multiple}); `Dust': dust absorption features (see
Section~\ref{ss:dust}); `Disk': the circumnuclear morphology has a high
ellipticity, suggesting the possible presence of an edge-on nuclear disk (see
Section~\ref{ss:stardisks}); `Nuc': bright point-like nucleus, most likely
associated with an AGN (see Section~\ref{ss:pointnuclei}); `Hollow':
depression in the surface brightness in the very nucleus (see
Section~\ref{ss:hollow}).}
\end{deluxetable}


\begin{deluxetable}{lc}
\tabletypesize{\tiny}
\tablecaption{DUST FEATURES\label{t:dust}}
\tablewidth{0pt}
\tablehead{
\colhead{BCG host cluster} & \colhead{Dust morphology}
}
\startdata
Abell 147 & D \\
Abell 160 & F \\
Abell 189 & D \\
Abell 193 & D, F \\
Abell 262 & F, P \\
Abell 397 & P \\
Abell 419 & P \\
Abell 496 & F \\
Abell 569 & D \\
Abell 671 & P \\
Abell 1060 & F, P \\
Abell 1308 & D \\
Abell 1836 & D \\
Abell 1983 & F \\
Abell 2052 & P \\
Abell 2247 & F \\
Abell 2593 & D \\
Abell 2634 & R \\
Abell 2657 & F, S \\
Abell 2666 & D \\
Abell 3526 & F, S \\
Abell 3559 & D \\
Abell 3565 & D \\
Abell 3676 & F \\
Abell 3677 & F, R \\
Abell 3698 & F, P, S \\
Abell 3733 & F, P \\
Abell 3744 & D, R \\
Abell 3747 & F, S \\
Abell 4049 & R \\
Abell 4059 & P \\ 
\enddata
\tablecomments{{\tiny
The table lists those clusters for which we found evidence for dust
absorption in the BCG. Column~(1) lists the identification of the
cluster in the catalogs of \citet{abell58} and \citet{abell89}.
Column~(2) classifies the morphology of the dust absorption: `D' =
dust disk in the nucleus; `F' = dust filaments; `P' = dust patches;
`R' = dust ring around the nucleus; `S' = dust spiral. Examples of
these morphologies are shown in Figure~\ref{f:dust}. Multiple
morphologies are listed for those galaxies for which no unambiguous
classification could be made.}}
\end{deluxetable}


\begin{deluxetable}{lccccccc}
\tabletypesize{\tiny}
\tablecaption{NUKER-LAW FIT PARAMETERS\label{t:nukerfits}}
\tablewidth{0pt}
\tablehead{
\colhead{BCG host cluster} & \colhead{$r_{\rm b}$} & \colhead{$I_0$} & 
\colhead{$\tau$} & \colhead{$\beta$} & \colhead{$\gamma$} & 
\colhead{$\GG$} & \colhead{Profile}\\
\colhead{} & \colhead{(arcsec)} & \colhead{($I$-band mag arcsec$^{-2}$)} &
\colhead{} & \colhead{} & \colhead{} & \colhead{} & \colhead{}
}
\startdata
ABELL76\tablenotemark{a} & 0.28 & 15.54 &  1.22 & 1.33 &  -0.20 & -0.03 & $\cap$ \\
ABELL119  & 0.80 & 17.16 &  3.12 & 1.06 &   0.06 &  0.06 & $\cap$ \\
ABELL168  & 0.06 & 15.22 &  0.95 & 1.02 &  -0.48 &  0.19 & $\cap$ \\
ABELL189  & 0.65 ($<$0.05) & 16.16 &  9.85 & 1.22 &   0.85 &  0.85 & $\backslash$ \\
ABELL193  & 0.39 & 15.61 &  2.82 & 1.48 &   0.23 &  0.23 & $\cap$ \\
ABELL194  & 0.70 & 15.50 &  1.53 & 1.47 &   0.08 &  0.10 & $\cap$ \\
ABELL195  & 0.49 & 16.09 &  2.12 & 1.41 &   0.09 &  0.10 & $\cap$ \\
ABELL260  & 0.75 & 16.31 &  3.35 & 1.29 &  -0.01 & -0.01 & $\cap$ \\
ABELL261  & 0.98 ($<$0.05) & 16.82 & 10.00 & 1.47 &   0.76 &  0.76 & $\backslash$ \\
ABELL295  & 0.63 & 16.53 &  3.81 & 1.24 &   0.13 &  0.13 & $\cap$ \\
ABELL347\tablenotemark{a} & 0.39 & 15.53 &  4.24 & 0.91 &  -0.03 & -0.03 & $\cap$ \\
ABELL376  & 0.68 & 16.96 &  2.38 & 1.28 &   0.19 &  0.20 & $\cap$ \\
ABELL397  & 1.06 & 16.62 &  2.69 & 1.50 &   0.07 &  0.07 & $\cap$ \\
ABELL419  & 0.47 ($<$0.1) & 16.37 &  0.60 & 1.64 &   0.33 &  0.60 & $\backslash$ \\
ABELL496  & 0.73 & 16.80 &  2.24 & 1.01 &   0.10 &  0.10 & $\cap$ \\
ABELL533  & 0.31 & 15.90 &  1.81 & 1.31 &   0.06 &  0.10 & $\cap$ \\
ABELL548  & 0.40 & 16.10 &  0.96 & 1.38 &   0.00 &  0.16 & $\cap$ \\
ABELL634\tablenotemark{a}  & 0.23 & 15.70 &  2.97 & 0.88 &  -0.05 & -0.04 & $\cap$ \\
ABELL779  & 1.04 & 15.97 &  1.69 & 1.44 &   0.02 &  0.03 & $\cap$ \\
ABELL912  & 0.21 ($<$0.08) & 15.38 &  1.71 & 1.34 &   0.48 &  0.55 & $\backslash$ \\
ABELL999  & 0.62 & 16.10 &  1.06 & 1.65 &  -0.06 &  0.05 & $\cap$ \\
ABELL1016 & 0.23 & 15.01 &  3.86 & 1.16 &   0.24 &  0.25 & $\cap$ \\
ABELL1142 & 0.19 & 14.46 &  9.33 & 1.31 &   0.12 &  0.12 & $\cap$ \\
ABELL1177 & 0.57 & 16.24 &  1.80 & 1.31 &   0.13 &  0.14 & $\cap$ \\
ABELL1228 & 0.27 ($<$0.08) & 15.21 &  1.43 & 1.41 &   0.53 &  0.60 & $\backslash$ \\
ABELL1314 & 1.19 & 16.66 &  2.34 & 1.60 &   0.17 &  0.17 & $\cap$ \\
ABELL1367 & 0.93 & 16.20 &  2.18 & 1.27 &   0.11 &  0.12 & $\cap$ \\
ABELL1631 & 0.20 & 15.25 &  1.52 & 1.29 &  -0.03 &  0.12 & $\cap$ \\
ABELL1656 & 2.10 & 16.63 &  2.11 & 1.46 &   0.03 &  0.03 & $\cap$ \\
ABELL1983 & 0.10 & 14.39 &  0.28 & 2.63 &  -1.37 &  0.44 & i \\
ABELL2040 & 0.39 & 16.40 &  1.71 & 1.39 &   0.16 &  0.19 & $\cap$ \\
ABELL2147 & 2.21 & 18.07 &  1.56 & 1.37 &   0.18 &  0.18 & $\cap$ \\
ABELL2162 & 1.04 & 16.35 &  1.39 & 1.78 &  -0.01 &  0.02 & $\cap$ \\
ABELL2197 & 0.78 & 15.88 &  0.60 & 1.86 &  -0.33 &  0.02 & $\cap$ \\
ABELL2247 & 1.99 ($<$0.03) & 17.99 &  1.85 & 1.42 &   0.85 &  0.85 & $\backslash$ \\
ABELL2572\tablenotemark{a} & 0.18 & 14.95 &  1.67 & 1.12 &  0.10 &  0.21 & $\cap$ \\
ABELL2589 & 0.25 & 15.79 &  3.38 & 1.10 &   0.04 &  0.05 & $\cap$ \\
ABELL2877 & 0.98 & 15.64 &  1.33 & 1.58 &  -0.02 &  0.01 & $\cap$ \\
ABELL3144 & 0.39 & 15.79 &  1.77 & 1.64 &   0.07 &  0.11 & $\cap$ \\
ABELL3193 & 0.26 & 15.30 &  1.35 & 1.37 &   0.08 &  0.20 & $\cap$ \\
ABELL3376 & 1.95 & 17.70 &  3.15 & 1.50 &   0.05 &  0.05 & $\cap$ \\
ABELL3395 & 0.37 & 16.69 &  2.43 & 0.98 &   0.03 &  0.04 & $\cap$ \\
ABELL3526 & 1.40 & 16.36 &  6.63 & 0.86 &   0.10 &  0.10 & $\cap$ \\
ABELL3528 & 0.61 & 16.41 &  2.49 & 1.35 &   0.18 &  0.18 & $\cap$ \\
ABELL3532 & 0.43 & 16.28 &  3.15 & 1.30 &   0.18 &  0.18 & $\cap$ \\
ABELL3554 & 0.52 & 17.02 &  2.74 & 1.10 &   0.04 &  0.04 & $\cap$ \\
ABELL3556 & 0.47 & 15.76 &  2.21 & 1.34 &   0.09 &  0.10 & $\cap$ \\
ABELL3558 & 1.73 & 17.97 &  2.08 & 1.10 &   0.05 &  0.05 & $\cap$ \\
ABELL3562 & 1.15 & 17.73 &  1.21 & 1.32 &   0.00 &  0.03 & $\cap$ \\
ABELL3564 & 0.24 & 15.56 &  1.34 & 1.38 &   0.05 &  0.20 & $\cap$ \\
ABELL3570 & 0.26 & 15.16 &  1.26 & 1.50 &   0.00 &  0.17 & $\cap$ \\
ABELL3571 & 1.15 & 17.88 &  2.85 & 0.75 &   0.02 &  0.02 & $\cap$ \\
ABELL3574 & 0.76 & 15.57 &  2.51 & 1.30 &  -0.02 & -0.02 & $\cap$ \\
ABELL3656 & 0.99 & 15.74 &  2.08 & 1.46 &   0.09 &  0.09 & $\cap$ \\
ABELL3677 & 0.27 & 15.49 &  1.28 & 1.63 &  -0.04 &  0.13 & $\cap$ \\
ABELL3716 & 0.44 & 16.62 &  2.42 & 1.10 &  -0.03 & -0.02 & $\cap$ \\
ABELL3736 & 0.92 & 16.89 &  1.37 & 1.33 &   0.11 &  0.13 & $\cap$ \\
ABELL3742\tablenotemark{a} & 0.22 & 14.21 &  2.22 & 1.05 &  0.08 &  0.11 & $\cap$ \\
ABELL3747\tablenotemark{a} & 0.18 & 14.64 &  2.10 & 1.16 &  -0.02 & 0.05  & $\cap$ \\
ABELL4038 & 0.33 & 15.37 &  1.16 & 1.44 &   0.03 &  0.17 & $\cap$ \\
\enddata	  
\tablecomments{Column~(1) lists the identification of the BCG host
cluster in the catalogs of \citet{abell58} and \citet{abell89}. The table includes
only the 60 sample galaxies for which the major axis surface
brightness profile of the BCG could be reliably determined from the
data, as described in Section~\ref{ss:sbanalysis}. Columns~(2)--(6)
list the parameters of the best Nuker-law (eq.~[\ref{nukerlaw}]) fit
to the brightness profile. The scale brightness $I_0$ includes
corrections for Galactic foreground extinction and bandshift
(K-correction). The quantity $\GG$ in column~(7) is the power-law
slope $-{\rm d}\,\log I / {\rm d}\,\log r$ of the
Nuker-law fit at 0\farcs 05 from the galaxy center. The type of the
surface brightness profile is given in column~(8): `$\cap$' indicates
a ``core'' profile, ``$\backslash$'' indicates a ``power-law'' profile, and
``i'' indicates an ``intermediate-slope'' profile. Fits to the surface 
brightness profile were generally performed out to $r$ = 10\arcsec,
as described in Section~\ref{ss:paramfits}. For those galaxies for
which the name in Column (1) is followed by an ``${\rm a}$,'' the fit was
done only out to $r$ = 2\arcsec. For the power-law and intermediate-slope
galaxies, we list in parenthesis in Column (2) an upper limit to any true 
``break,'' determined as described in Section~\ref{ss:corepower}.}
\end{deluxetable}


\begin{deluxetable}{lcccccccccc}
\tabletypesize{\tiny}
\rotate
\tablecaption{PROPERTIES OF BCGS AND THEIR HOST CLUSTERS\label{t:properties}}
\tablewidth{0pt}
\tablehead{
\colhead{Cluster} & \colhead{BM type} & \colhead{$M_{R}$} & \colhead{$\alpha$} & \colhead{$\Delta$} &
\colhead{$B-R$} & \colhead{Richness} & \colhead{Sep.} & \colhead{Vel. Diff.} &
\colhead{Vel. Disp.} & \colhead{$L_{\rm X}$} \\ 
\colhead{} & \colhead{} & \colhead{($10~\kpc$)} & \colhead{} & \colhead{} & \colhead{} & 
\colhead{} &   \colhead{(Mpc)} & \colhead{(km~s$^{-1}$)} & 
\colhead{(km~s$^{-1}$)} & \colhead{($10^{44}$~ergs~s$^{-1}$)}  
}
\startdata
  76 & II-III  &  -22.518 & 0.555 & -0.026 &  1.674 &   42 &    0.225 &      0 & \nodata&  0.453 \\
 119 & II-III  &  -22.749 & 0.766 & -0.092 &  1.515 &   69 &    0.088 &    -35 & 753.0& 2.310 \\
 147 & III  &  -22.533 & 0.427 & -0.260 &  1.486 &   32 &    0.390 &    116 & \nodata& \nodata \\
 160 & III  &  -22.184 & 0.752 &  0.469 &  1.546 &   34 &    0.140 &      0 & \nodata& \nodata \\
 168 & II-III &  -22.573 & 0.576 & -0.054 &  1.495 &   89 &   0.583  &    -7  &458.2 & \nodata \\
 189 & III  &  -21.932 & 0.337 &  0.134 &  1.489 &   50 &   0.159  &   310  &\nodata & \nodata \\
 193 & II  &  -22.644 & 0.684 & -0.023 &  1.515 &   58 &   0.000  &   272  &\nodata & \nodata \\
 194 & II  &  -22.497 & 0.626 &  0.077 &  1.483 &   37 &   0.223  &   -37  & 422.9& 0.133 \\
 195 & II  &  -22.430 & 0.469 & -0.075 &  1.431 &   32 &   0.062  &     0  &\nodata & 0.142 \\
 260 & II  &  -22.724 & 0.588 & -0.191 &  1.509 &   51 &   0.616  &  -300  & 553.6& \nodata \\
 261 & I  &  -22.575 & 0.570 & -0.065 &  1.499 &   63 &   0.062  &     0  & \nodata & \nodata \\
 262 & III  &  -22.189 & 0.810 &  0.471 &  1.545 &   40 &   0.031  &  -100  &506.1 & 0.579 \\
 295 & II  &  -22.532 & 0.563 & -0.030 &  1.505 &   51 &   0.188  &   -73  &\nodata & \nodata \\
 347 & II-III  &  -22.352 & 0.601 &  0.197 &  1.495 &   32 &   0.108  &  -404  & 690.0& \nodata \\
 376 & I-II  &  -22.553 & 0.696 &  0.076 &  1.522 &   36 &   0.217  &     0  &\nodata & 1.123 \\
 397 & III  &  -22.542 & 0.582 & -0.016 &  1.490 &   35 &   0.269  &   310  &\nodata & 0.081 \\
 419 & \nodata  &  -21.851 & 0.333 &  0.206 &  1.442 &   32 &   0.125  &     0  & \nodata& 0.138 \\
 496 & I  &  -22.676 & 0.803 & -0.016 &  1.541 &   50 &   0.000  &     0  & 687.0&  3.188\\
 533 & \nodata  &  -22.471 & 0.506 & -0.054 &  1.451 &   31 &   0.280  &   181  & \nodata& \nodata \\
 548 & III  &  -22.561 & 0.498 & -0.156 &  1.499 &   79 &   1.463  &  -101  & 502.6& \nodata \\
 569 & II  &  -22.418 & 0.486 & -0.032 &  1.431 &   36 &   0.031  &   -25  & \nodata& 0.055 \\
 634 & III  &  -22.258 & 0.498 &  0.147 &  1.575 &   40 &   0.556  &     0  & 328.0& \nodata \\
 671 & II-III  &  -22.965 & 0.713 & -0.327 &  1.524 &   38 &   0.000  &  -250  & \nodata& 0.804 \\
 779 & I-II  &  -22.858 & 0.594 & -0.318 &  1.565 &   32 &   0.054  &    70  &481.5 & 0.072 \\
 912 & \nodata  &  -21.948 & 0.419 &  0.309 &  1.497 &   36 &   0.062  &     0  & \nodata& \nodata \\
 999 & II-III  &  -22.267 & 0.441 &  0.034 &  1.522 &   33 &   0.044  &   171  & 233.1& 0.033 \\
1016 & \nodata  &  -22.048 & 0.430 &  0.231 &  1.532 &   37 &   0.077  &    37  & 225.9& \nodata \\
1060 & III  &  -22.275 & 0.818 &  0.385 &  1.535 &   50 &   0.031  &   -16  & 597.9& 0.457 \\
1142 & II-III  &  -22.295 & 0.545 &  0.182 &  1.521 &   35 &   0.077  &  -383  & 952.6& 0.176 \\
1177 & I  &  -22.453 & 0.724 &  0.190 &  1.525 &   32 &   0.188  &     0  &181.5 & \nodata \\
1228 & II-III  &  -22.133 & 0.431 &  0.149 &  1.476 &   50 &   0.062  &  -299  & 1129.5& \nodata \\
1308 & II-III   &  -22.768 & 0.554 & -0.277 &  1.485 &   37 &   0.217  &     0  &\nodata & \nodata \\
1314 & III  &  -22.461 & 0.583 &  0.066 &  1.546 &   44 &   0.044  &   154  & \nodata& 0.462 \\
1367 & II-III  &  -22.496 & 0.518 & -0.058 &  1.545 &  117 &   0.212  &  -233  &818.8  & 1.296 \\
1631 & I  &  -22.566 & 0.649 &  0.028 &  1.538 &   34 &   0.428  &   170  & 682.0& 0.522 \\
1656 & II  &  -22.957 & 0.590 & -0.421 &  1.531 &  106 &   0.108  &  -464  &656.0 & 6.798 \\
1836 & II  &  -22.622 & 0.577 & -0.102 &  1.507 &   41 &   0.062  &   -37  &\nodata & \nodata \\
1983 & III  &  -22.226 & 0.325 & -0.191 &  1.552 &   51 &   0.446  &   139  & 379.9& 0.352 \\
2040 & III  &  -22.071 & 0.750 &  0.581 &  1.533 &   52 &   0.000  &   -18  & \nodata& 0.344 \\
2052 & I-II  &  -22.479 & 0.879 &  0.165 &  1.546 &   41 &   0.000  &  -244  &551.9 & 1.992 \\
2147 & III   &  -22.262 & 0.641 &  0.326 &  1.576 &   52 &   0.140  &  -128  &875.0 & 1.887 \\
2162 & II-III  &  -22.475 & 0.503 & -0.062 &  0.000 &   37 &   0.117  &   -82  &\nodata & \nodata \\
2197 & III  &  -22.887 & 0.586 & -0.356 &  0.000 &   73 &   0.596  &  -242  &602.7 & 0.076 \\
2247 & III  &  -22.260 & 0.407 & -0.029 &  0.000 &   35 &   0.527  &   375  &\nodata & \nodata \\
2572 & III  &  -22.586 & 0.530 & -0.130 &  1.655 &   32 &   0.740  &   589  & \nodata& \nodata \\
2589 & I  &  -22.420 & 0.781 &  0.239 &  1.407 &   40 &   0.108  &  -253  &734.6 & \nodata \\
2593 & II  &  -22.498 & 0.800 &  0.162 &  1.481 &   42 &   0.153  &    24  & \nodata& 1.037 \\
2634 & II  &  -22.748 & 0.650 & -0.153 &  1.519 &   52 &   0.077  &   -13  &962.9 & 0.693 \\
2657 & III  &  -21.998 & 0.350 &  0.100 &  0.000 &   51 &   0.337  &   354  & \nodata& 1.697 \\
2666 & I  &  -22.768 & 0.549 & -0.285 &  1.464 &   34 &   0.000  &    65  &838.2 & \nodata \\
2877 & I  &  -23.284 & 0.612 & -0.725 &  1.545 &   30 &   0.070  &   -48  & \nodata& 0.381 \\
3144 & I-II  &  -22.171 & 0.451 &  0.150 &  1.484 &   54 &   0.000  &  -427  & \nodata & \nodata \\
3193 & I  &  -22.447 & 0.498 & -0.041 &  1.479 &   41 &   0.000  &  -342  & 787.6& \nodata \\
3376 & I  &  -22.697 & 0.615 & -0.133 &  1.488 &   42 &   0.000  &   -94  &680.9 & 1.722 \\
3395 & II  &  -22.490 & 0.795 &  0.170 &  1.526 &   54 &   0.153  &  -192  &\nodata & 2.482 \\
3526 & I-II  &  -22.885 & 0.731 & -0.239 &  1.482 &   33 &   0.000  &  -409  & 619.3& 1.645 \\
3528 & II   &  -22.924 & 0.667 & -0.315 &  0.000 &   70 &   0.088  &   108  & \nodata& 3.136 \\
3532 & II-III  &  -22.668 & 0.721 & -0.027 &  0.000 &   36 &   0.000  &   -14  &739.4 & 1.708 \\
3554 & I-II  &  -22.386 & 0.701 &  0.245 &  1.883 &   59 &   0.000  &     0  &\nodata & \nodata \\
3556 & I  &  -22.948 & 0.565 & -0.443 &  1.485 &   49 &   0.000  &   -42  &\nodata & \nodata \\
3558 & I  &  -23.052 & 0.893 & -0.413 &  1.482 &  226 &   0.000  &  -203  &952.3 & \nodata \\
3559 & I  &  -22.906 & 0.627 & -0.331 &  1.520 &  141 &   0.088  &  -108  &\nodata & \nodata \\
3562 & I  &  -22.561 & 0.695 &  0.067 &  1.495 &  129 &   0.000  &     0  &\nodata & 0.680 \\
3564 & II  &  -22.291 & 0.435 & -0.002 &  1.477 &   53 &   0.512  &  -225  &\nodata & \nodata \\
3565 & I  &  -22.547 & 0.525 & -0.099 &  1.358 &   64 &   0.000  &   -72  & \nodata& 0.011 \\
3570 & I-II  &  -22.168 & 0.324 & -0.135 &  0.000 &   31 &   0.234  &   221  & \nodata& \nodata \\
3571 & I  &  -22.881 & 1.101 & -0.462 &  1.516 &  126 &   0.000  &  -234  & \nodata& \nodata \\
3574 & I  &  -22.471 & 0.744 &  0.180 &  1.524 &   31 &   0.031  &  -135  & 639.9& \nodata \\
3656 & I-II  &  -22.679 & 0.617 & -0.114 &  1.387 &   35 &   0.094  &   198  &\nodata & \nodata \\
3676 & II-III  &  -22.494 & 0.473 & -0.133 &  1.362 &   33 &   0.455  &     0  &\nodata & \nodata \\
3677 & I  &  -21.910 & 0.414 &  0.336 &  1.499 &   60 &   0.000  &     0  & \nodata& \nodata \\
3698 & I-II  &  -22.067 & 0.370 &  0.079 &  1.518 &   71 &   0.031  &  -241  & \nodata& \nodata \\
3716 & I-II  &  -22.559 & 0.709 &  0.077 &  1.481 &   66 &   0.293  &   474  &736.5 & \nodata \\
3733 & I-II  &  -22.128 & 0.643 &  0.462 &  1.561 &   59 &   0.000  &     2  &\nodata & \nodata \\
3736 & III  &  -22.966 & 0.678 & -0.349 &  1.499 &   35 &   0.125  &     0  &\nodata & \nodata \\
3742 & II-III  &  -21.985 & 0.424 &  0.281 &  1.509 &   35 &   0.215  &   -78  &\nodata & \nodata \\
3744 & II-III  &  -22.514 & 0.423 & -0.249 &  1.568 &   70 &   0.062  &   -68  &\nodata & \nodata \\
3747 & I-II  &  -22.268 & 0.449 &  0.048 &  1.513 &   44 &   0.044  &     4  & \nodata& \nodata \\
4038 & III  &  -22.379 & 0.454 & -0.052 &  1.459 &  117 &   0.108  &  -284  & 272.9 & \nodata \\
4049 & III  &  -22.529 & 0.516 & -0.095 &  1.518 &   39 &   0.000  &  -376. &547.3 & \nodata \\
4059 & I  &  -22.928 & 0.899 & -0.292 &  1.527 &   66 &   0.348  &   -34. & \nodata& \nodata \\
\enddata
\tablecomments{\tiny Column~(1) lists the identification of the cluster in 
the catalogs of \citet{abell58} and \citet{abell89}. Column~(2) lists the
Bautz-Morgan cluster morphology classification
\citep{leir77,bautz70,bautz72,cor74,sand76,kris78,white78,abell89}. Column~(3) 
lists the absolute $R$-band metric magnitude $M_{R}$ ($10 \kpc)$ of the BCG
inside an aperture of $10 \kpc$ radius (calculated from the values in
Table~\ref{t:sample}). Column~(4) lists the parameter $\alpha$, which
measures the logarithmic slope of the metric luminosity as a function
of radius, determined at a physical radius of 10 kpc. Column~(5) lists
the residual $\Delta$ between the observed metric luminosity and that
predicted by the BCG standard-candle relation between metric
luminosity and $\alpha$. Column~(6) lists the $B-R$ color at the metric
radius after correcting for Galactic extinction and
K-dimming. Column~(7) lists the Abell richness count (number of
galaxies with magnitudes between $m_{3}$ and $m_{3+2}$ within an Abell
radius of the cluster center, where $m_{3}$ is the magnitude of the
third brightest cluster galaxy). Column~(8) lists the projected
separation of the BCG from the cluster center. Column~(9) lists the
difference between the line-of-sight velocity of the BCG and the
systemic velocity of the cluster. Column~(10) lists the available velocity 
dispersion for clusters which have measured redshifts for at least 20 member 
galaxies (M. Postman, private communication).
Column~(11) lists the X-ray luminosities from the compilation of
\citet{jones99}. All data are based on \citet{pos95}, unless 
otherwise noted.}
\end{deluxetable}



\end{document}